\documentclass[aps,prc,twocolumn,superscriptaddress, floatfix]{revtex4-1}

\usepackage[utf8]{inputenc}
\usepackage[english]{babel}
\usepackage{bm}
\usepackage{graphicx}
\usepackage{amsmath}
\usepackage{amssymb}
\usepackage{booktabs}
\usepackage{enumitem}
\usepackage{color}
\usepackage{multirow}
\usepackage[colorlinks=true,linkcolor=blue,allcolors=blue]{hyperref} % for links 
\usepackage[caption=false]{subfig}
\usepackage{xcolor}

\usepackage{verbatimbox}

\usepackage [autostyle, english = american]{csquotes}
\MakeOuterQuote{"}

\usepackage{mathtools} 	%% because math++
\usepackage{comment} 	%% block comments 
\usepackage{multirow}	%% for longer eq.
\usepackage{gensymb}
\usepackage[binary-units=true]{siunitx}  %% because of \mu vs \si{\micro}
\DeclareSIUnit\sample{S}
\DeclareSIUnit\bits{bits}

\usepackage{lineno} 	%% for line numbers
\newcommand*\patchAmsMathEnvironmentForLineno[1]{%
  \expandafter\let\csname old#1\expandafter\endcsname\csname #1\endcsname
  \expandafter\let\csname oldend#1\expandafter\endcsname\csname end#1\endcsname
  \renewenvironment{#1}%
     {\linenomath\csname old#1\endcsname}%
     {\csname oldend#1\endcsname\endlinenomath}}% 
\newcommand*\patchBothAmsMathEnvironmentsForLineno[1]{%
  \patchAmsMathEnvironmentForLineno{#1}%
  \patchAmsMathEnvironmentForLineno{#1*}}%
\AtBeginDocument{%
\patchBothAmsMathEnvironmentsForLineno{equation}%
\patchBothAmsMathEnvironmentsForLineno{align}%
\patchBothAmsMathEnvironmentsForLineno{flalign}%
\patchBothAmsMathEnvironmentsForLineno{alignat}%
\patchBothAmsMathEnvironmentsForLineno{gather}%
\patchBothAmsMathEnvironmentsForLineno{multline}%
}
\usepackage{xspace} 	%% for commands that don't eat spaces afterwards
\newcommand{\PI}{Phase I\xspace}
\newcommand{\PII}{Phase II\xspace}
\newcommand{\PIIa}{Phase IIa\xspace}
\newcommand{\PIIb}{Phase IIb\xspace}

\newcommand{\axcoup}{$g_{\alpha\gamma\gamma}$\xspace}
\newcommand{\gagg}{\axcoup}
\newcommand{\ggam}{$g_{\gamma}$\xspace}
\newcommand{\gKSVZ}{$g^{\textrm{KSVZ}}_{\gamma}$\xspace}
\newcommand{\AxionLineWidth}{$\sim$\SI{5}{\kHz}\xspace}
\newcommand{\DarkMatterDen}{\SI{0.45}{\GeV/cm^{3}}}
\newcommand{\SNR}{$\Sigma$\xspace}
\newcommand{\detune}{$\delta_{\nu}$\xspace}

\newcommand{\chirfl}{$\left| \chi_{rfl} \right|^{2}$\xspace}
\newcommand{\axpwr}{$P_\text{ax}$\xspace}
\newcommand{\SQZ}{SQ\xspace}
\newcommand{\AMP}{AMP\xspace}
\newcommand{\TMC}{$T_\text{mc}$\xspace}
\newcommand{\Nsys}{$N_\text{sys}$\xspace}

\newcommand{\PIIbfwidth}{\SI{64}{\MHz}\xspace}
\newcommand{\PIIabfwidth}{\SI{137}{\MHz}\xspace}

\newcommand{\PIfrange}{\SIrange[range-phrase=--,range-units=single]{5.6}{5.8}{GHz}\xspace}

\newcommand{\PIIafrange}{4.100--\SI{4.140}{\GHz} and 4.145--\SI{4.178}{\GHz}\xspace}
\newcommand{\PIIbfrange}{4.459-\SI{4.523}{\GHz}\xspace}

\newcommand{\PImrange}{\SIrange[range-phrase=--,range-units=single]{23.15}{24.0}{\micro\eV\per c\squared}\xspace}
\newcommand{\PIIamrange}{16.96--17.12 and 17.14--\SI{17.28}{\micro\eV\per c^{2}}\xspace}
\newcommand{\PIIbmrange}{18.44--\SI{18.71}{\micro\eV\per c^{2}}\xspace}

\newcommand{\PIIaBayAgg}{1.95$\times$\gKSVZ\xspace}
\newcommand{\PIIbBayAgg}{2.06$\times$\gKSVZ\xspace}

\newcommand{\daqenhance}{$\sim$1.6\xspace}

\newcommand{\PIIbDeadPercent}{\SI{18}{\percent}\xspace}
\newcommand{\PIIaDeadPercent}{\SI{52}{\percent}\xspace}

\newcommand{\basetempNoErr}{\SI{61}{mK}\xspace} % Base temp
\newcommand{\PIIaVTS}{\SI{333}{m\kelvin}\xspace}

\newcommand{\PIIbVTS}{\SI{290}{m\kelvin}\xspace}
\newcommand{\magfield}{\SI{8}{T}\xspace} % Nominal B-Field
\newcommand{\MagBoreSize}{\SI{12.7}{cm}\xspace}

\newcommand{\CavityFreqRange}{3.6--\SI{5.8}{GHz}\xspace}
\newcommand{\cavlength}{\SI{25.4}{\cm}\xspace} % cavity height
\newcommand{\cavdiameter}{\SI{10.2}{\cm}\xspace} % cavity diameter
\newcommand{\cavaxionvolumeNoErr}{\SI{1.545}{\litre}\xspace}
\newcommand{\roddiameter}{\SI{5.1}{\cm}\xspace} % rod diameter
\newcommand{\rodgap}{$\sim$\SI{250}{\micro\metre}\xspace} % gap size
\newcommand{\CuStemDiameter}{\SI{3.175}{m\metre}\xspace} % stem diameter
\newcommand{\AxleDiameter}{\SI{6.35}{m\metre}\xspace}

\newcommand{\jpaheight}{$\sim$\SI{60}{\cm}\xspace}

\newcommand{\Nadd}{$N_{a}$\xspace}
\newcommand{\SNRThreshold}{3.468$\sigma$\xspace}
\newcommand{\SNRTarget}{5.1$\sigma$\xspace}

\newcommand{\PIIaEta}{0.63\xspace}
\newcommand{\PIIbEta}{0.60\xspace}

\makeatletter
\renewcommand\tableofcontents{%
    \@starttoc{toc}%
}
\makeatother

\newif\ifshowchanges
\showchangesfalse
\newcommand{\vo}[1]{}

\ifshowchanges
\usepackage{soul}
\soulregister\cite7
\soulregister\ref7
	\renewcommand{\vo}[1]{{\textcolor{red}{\st{#1}}}}
    
\fi

\usepackage{fixme}
\fxsetup{status=draft,theme=color, layout=inline}
\definecolor{fxnote}{rgb}{0.8000,0.0000,0.0000}

\begin{document}

\title{New Results from HAYSTAC's Phase II Operation with a Squeezed State Receiver}
\collaboration{HAYSTAC Collaboration}

\newcommand{\Location}{\affiliation{Location}}
\newcommand{\Yale}{\affiliation{Department of Physics, Yale University, New Haven, Connecticut 06520, USA}}
\newcommand{\YaleApplied}{\affiliation{Department of Applied Physics, Yale University, New Haven, Connecticut 06520, USA}}
\newcommand{\WrightLab}{\affiliation{Wright Laboratory, Department of Physics, Yale University, New Haven, Connecticut 06520, USA}}
\newcommand{\Cal}{\affiliation{Department of Nuclear Engineering, University of California Berkeley, California 94720, USA}}
\newcommand{\JILA}{\affiliation{JILA, National Institute of Standards and Technology and the University of Colorado, Boulder, Colorado 80309, USA}}
\newcommand{\Colorado}{\affiliation{Department of Physics, University of Colorado, Boulder, Colorado 80309, USA}}
\newcommand{\Hopkins}{\affiliation{Department of Physics and Astronomy, The Johns Hopkins University, 3400 North Charles Street Baltimore, MD, 21211}}
\newcommand{\NIST}{\affiliation{National Institute of Standards and Technology, Boulder, Colorado 80305, USA}}

\author{M.J.~Jewell}\Yale\WrightLab
\author{A.F.~Leder}\Cal
\author{K.M.~Backes}\altaffiliation{now at The MITRE Corporation}\Yale\WrightLab
\author{Xiran~Bai}\Yale\WrightLab
\author{K.~van~Bibber}\Cal
\author{B.M.~Brubaker}\Colorado
\author{S.B.~Cahn}\Yale\WrightLab
\author{A.~Droster}\Cal
\author{Maryam H.~Esmat}\Hopkins
\author{Sumita~Ghosh}\YaleApplied\WrightLab
\author{Eleanor~Graham}\Yale\WrightLab
\author{Gene C.~Hilton}\NIST
\author{H.~Jackson}\Cal
\author{Claire~Laffan}\Yale\WrightLab
\author{S.K.~Lamoreaux}\Yale\WrightLab
\author{K.W.~Lehnert}\Colorado
\author{S.M.~Lewis}\altaffiliation{now at FNAL}\Cal
\author{M.~Malnou}\Colorado
\author{R.H.~Maruyama}\Yale\WrightLab
\author{D.A.~Palken}\altaffiliation{now at US Senate Office of Sen. Hickenlooper}\Colorado
\author{N.M.~Rapidis}\altaffiliation{now at Stanford}\Cal
\author{E.P.~Ruddy}\Colorado
\author{M.~Simanovskaia}\altaffiliation{now at Stanford}\Cal
\author{Sukhman~Singh}\Yale\WrightLab
\author{D.H.~Speller}\Hopkins
\author{Leila R.~Vale}\NIST
\author{H.~Wang}\altaffiliation{now at Stanford}\Yale\WrightLab
\author{Yuqi~Zhu}\Yale\WrightLab
\date{\today}

\begin{abstract}

A search for dark matter axions with masses $>$\SI{10}{\micro\eV\per c\squared} has been performed using the HAYSTAC experiment's squeezed state receiver to achieve sub-quantum limited noise.  This report includes details of the design and operation of the experiment previously used to search for axions in the mass ranges \PIIamrange(\PIIafrange) as well as upgrades to facilitate an extended search at higher masses.  These upgrades include improvements to the data acquisition routine which have reduced the effective dead time by a factor of 5, allowing for the new region to be scanned \daqenhance times faster with comparable sensitivity.  No statistically significant evidence of an axion signal is found in the range \PIIbmrange(\PIIbfrange), leading to an aggregate upper limit exclusion at the \SI{90}{\percent} level on the axion-photon coupling of \PIIbBayAgg.       

\end{abstract}

\maketitle
%\tableofcontents
%\linenumbers

\section{Introduction}
\label{sec:intro}

Astrophysical and cosmological observations indicate that more than a quarter of the universe's mass-energy exists in the form of dark matter~\cite{planck_collaboration_planck_2020, bullet_cluster}. While there are a number of hypotheses for the nature of this dark matter, no candidate has been detected directly. One such candidate, the axion, was originally proposed as a solution to the strong $CP$ problem~\cite{peccei1977CP, peccei1977CP2} but also has the necessary characteristics to act as the dominant source of dark matter in the universe~\cite{weinberg1978boson, wilczek1978PT, preskill1983cosmology, Abbott1983bound, dine1983harmless}. Models of the axion do not explicitly predict its mass or coupling strength to matter, but for QCD axions all its couplings, g$_{aii}$, are proportional to its mass, m$_{a}$.  This includes its coupling to radiation, \gagg, which is relevant to this experiment. The resulting parameter space of the allowed masses and couplings is typically benchmarked by couplings from the Kim, Shifman, Vainshtein, and Zakharov~(KSVZ)~\cite{kim1979KSVZ, shifman1980KSVZ2} and the Dine, Fischler, Srednicki, and Zhitnitsky~(DFSZ)~\cite{dine1981DFSZ, zhit1980DFSZ2} models based on heavy quarks and multi-Higgs interactions respectively. The ability to solve two outstanding questions in physics, coupled with null results from searches for weakly interacting massive particles~(WIMPs), situate the axion as one of the most promising candidates for a new fundamental particle~\cite{bertone_2018_nature}.

Although the mass of the QCD axion is largely unknown, axions in the mass range $\sim$\SIrange[range-phrase=--,range-units=single]{1}{100}{\micro\eV\per c\squared} are favored by models in which the Peccei-Quinn~(PQ) symmetry breaking occurs after inflation~\cite{gorghetto2019mass,klaer2017dark,buschmann2020motivation}.  To date, the most sensitive probes for dark matter axions in this range are those using axion haloscopes~\cite{Kim:2022hmg, bartram2021search, RBF, UF, QUAX, ORGAN, backes2021quantum}.  These experiments exploit the axion's coupling to the pseudoscalar electromagnetic product $\textbf{E} \cdot \textbf{B}$ by employing a large static magnetic field $\textbf{B}$ to convert the oscillating axion field into a small electric field $\textbf{E}$ which oscillates at frequency $\nu_{a} \approx m_{a}c^{2}/h$ for axions of mass $m_{a}$~\cite{Sikivie:1983ip_halotheory,Sikivie:1985yu_halotheory}. To resonantly enhance the relatively small coupling strength, \gagg, expected for this interaction, the induced field is enclosed inside of a high quality factor cavity with a spatially overlaping cavity mode whose bandwidth encompasses $\nu_{a}$.  When this resonant condition is met, a small power excess at $\nu_{a}$ with spectral width given by the axion's virial velocity $v\sim$\SI{270}{k\meter\per\second} would be present.  Through the use of the haloscope technique, experiments have achieved ratios of the expected axion-induced signal power to the total experimental noise power~(SNR) which allow for probes of dark matter axions sensitive to QCD couplings. 

The main challenge of such an experiment arises from the fact that the neither the axion's mass nor coupling strength are \emph{a priori} known, requiring that a successful search maintain this high sensitivity for axions of many different masses.  Because the width of the cavity's resonant mode is small compared to the range of possible axion masses,  the defining metric of a haloscope is not the SNR for a single axion mass but rather the rate $R$ (in \si{\Hz\per\second}) at which axions of different masses can be searched for at a given SNR.  While modern haloscopes have achieved sensitivity to QCD axions over large mass ranges below $\sim$\SI{10}{\micro\eV\per c\squared} ~\cite{backes2021quantum,bartram2021search,Kim:2022hmg}, extending these searches to higher frequencies becomes increasingly challenging as a number of experimental parameters, most notably the cavity volume, scale poorly with frequency.  In order to maintain sensitivity to QCD axions at masses $\gtrsim$\SI{10}{\micro\eV\per c\squared}, improvements are needed to counter this drop in $R$. 

One such path is the reduction of the total noise of the system, which can in part be achieved through the use of low-noise amplifiers operating at cryogenic temperatures but is traditionally limited by quantum noise from zero-point vacuum fluctuations.  These fluctuations impose a standard quantum limit~(SQL) to the noise a haloscope can achieve with phase insensitive linear amplifiers. In recent years it has been demonstrated that circumvention of this barrier is possible by coupling the cavity to a squeezed state receiver~(SSR)~\cite{malnou2019squeezed} consisting of two Josephson parametric amplifiers~(JPAs)~\cite{yamamoto2008flux}.  Operation of these JPAs as phase sensitive amplifiers allows for the manipulation of squeezed vacuum states, in which the measurement uncertainty is moved into one quadrature of the field, resulting  in a reduction of noise below the SQL.  This allows for the cavity's measurement bandwidth to be substantially increased by strongly overcoupling the readout antenna which in turns gives an enhancement to the scan rate. This was first demonstrated in an axion search by the Haloscope At Yale Sensitive To Axion Cold dark matter (HAYSTAC)~\cite{backes2021quantum,Kelly_thesis, palken2020thesis}.  

In this work, details from HAYSTAC's quantum enhanced search for dark matter axions are presented in the context of the second phase of data taking with a focus on new results from continued operation of the SSR since the initial demonstration published in Ref.~\cite{backes2021quantum}.  A detailed overview of the experimental configuration is presented in Section~\ref{sec:overview} highlighting both changes made to the detector in order to integrate the SSR as well as improvements made since the initial SSR run to further improve the scan rate during the acquisition of new data.   Following this, results from HAYSTAC's dark matter search are presented in Section~\ref{sec:results} with focus on new data taken with the upgraded detector.

\section{Experimental Overview}
\label{sec:overview}

The HAYSTAC experiment began operation in 2016 and is now in its second phase of axion sensitive measurements. In \PI which ran from January~2016--July~2017, HAYSTAC took data with a single Josephson Parametric Amplifier (JPA) operated in phase-insensitive mode to scan the mass range of \PImrange(\PIfrange)~\cite{brubaker2017first, zhong2018results} achieving nearly quantum-limited noise for the first time in a haloscope experiment. In order to push beyond this quantum-limit, the receiver chain was modified in August~2018 to incorporate an SSR and HAYSTAC began a second phase of data taking refered to as \PII. The first iteration with this setup, \PIIa, operated from September~2019--April~2020 and searched for axions with masses between \PIIamrange(\PIIafrange) achieving noise below the quantum-limit for the first time in an axion experiment~\cite{backes2021quantum}. Following the conclusion of this search, a new set of JPAs optimized for frequencies $>$\SI{4.2}{\GHz} were installed.  This allowed for a second search, \PIIb, to be performed in the higher mass range between \PIIbmrange~(\PIIbfrange) between July~2021--November~2021. This also included upgrades to the data acquisition routine, increasing the effective livetime and in turn the scan rate of the experiment by a factor of \daqenhance.

\subsection{Principle of Squeezed State Receiver Enhancement }
\label{sec:ssr_det_principle}

\begin{figure}
    \centering
    \includegraphics[width=\columnwidth]{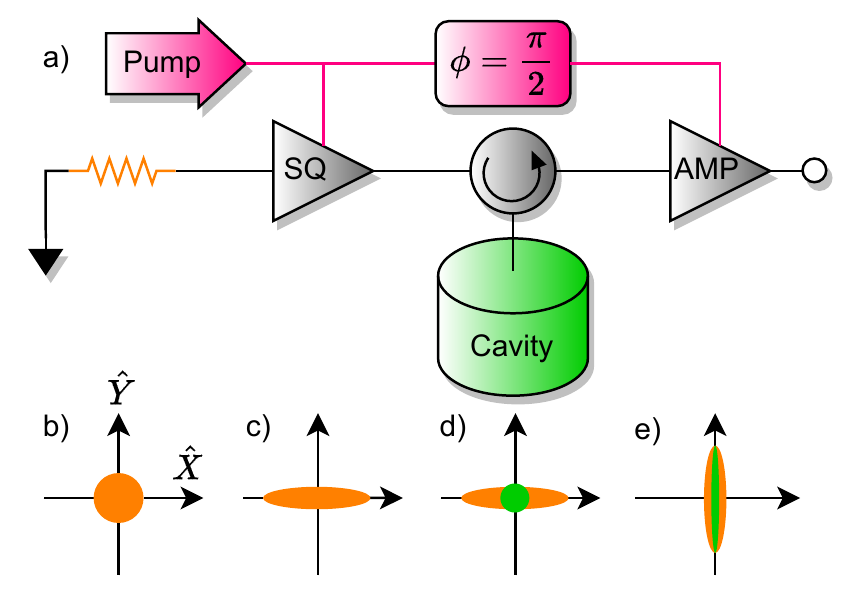}
    \caption{(a) Simplified illustration of the cavity coupled to the SSR outlining the main components and (b-e) how they transform the state in quadrature phase space. (b) An initially un-squeezed vacuum state (orange) sourced by a \SI{50}{\ohm} terminator is (c) sent to the \SQZ JPA which squeezes the $\hat{Y}$ quadrature. The state is then  (d) reflected off of the cavity thereby adding cavity noise (green) which would include any axion induced signal. (e) The state is then routed to the \AMP JPA, which is operated $\pi/2$ out of phase of the \SQZ JPA such that it amplifies the previously squeezed $\hat{Y}$ quadrature which contains both cavity noise as well as squeezed vacuum noise.}
    \label{fig:ssr_simple}
\end{figure}

The benefit of an SSR to a haloscope experiment was first described in detail in Ref.~\cite{malnou2019squeezed,Malnou:2017udw}.  The basic design of the SSR enhanced haloscope is shown in Fig.~\ref{fig:ssr_simple} where a microwave cavity converts axions into a detectable signal which is measured by a receiver chain consisting of two JPAs.  In such an experiment, the presence or absence of an axion is determined through precise measurements of the cavity's field, which is governed by the Hamiltonian
\begin{equation}
    \hat{H} = \frac{h\nu_{c}}{2}\left(\hat{X}^{2} + \hat{Y}^{2}\right)
\end{equation}
where $\hat{X}$ and $\hat{Y}$ are the non-commuting quadratures of the cavity field.  The precision by which the magnitude of this field can be measured is ultimately limited by the total noise of the system. If this noise is evenly distributed along both quadratures, any observation which attempts to measure the magnitude of both quadratures simultaneously is subject to the SQL for haloscopes which imposes a minimum required measurement uncertainty~\cite{Lamoreaux:2013koa}. Circumventing this limit is possible with the use of phase sensitive amplifiers, such as JPAs, which allow the distribution of noise to be manipulated in the quadrature phase space.  In the case of the SSR this is achieved by operating two JPAs in line with each other with the cavity coupled between them.  The first JPA, named the squeezer~(\SQZ), allows for the noise incident on the cavity to be prepared in a squeezed state, with the majority of the uncertainty pushed into a single quadrature.  The state is then reflected off of the cavity where it picks up noise from the cavity, such as an axion signal, which gets evenly distributed along both quadratures.  A measurement of the state is then made with the second JPA, the amplifier~(\AMP), which noiselessly amplifies the state along the previously squeezed quadrature, resulting in a reduction of the total noise of the system (\Nsys).

In this picture, the $\hat{X}$ and $\hat{Y}$ variables are functions of the frequency difference or detuning (\mbox{\detune}=($\nu_{a}-\nu_{c}$)) between the axion's frequency ($\nu_{a}$) and the cavity's frequency ($\nu_{c}$), equivalent to half the JPA's pump frequency ($\nu_{p}$), and are comprised of linear combinations of phasor components spaced equally around $\nu_{p}$ at $\pm$\detune.  The reduction to the total noise is understood by examining the three contributions to \Nsys in terms of these phasor variables:
\begin{equation}
    \mbox{\Nsys}(\mbox{\detune}) = N_{c}(\mbox{\detune}) + N_{r}(\mbox{\detune}) + N_{A}(\mbox{\detune})
    \label{eq:noise_components}
\end{equation}
expressed in units of power spectral density and assuming that the noise sources are statistically independent. The first term represents the Johnson-Nyquist noise sourced from the cavity's internal loss, $N_{c}$,  which peaks on cavity resonance, with the same Lorentzian lineshape as the cavity's transmission response.  This noise would also contain the axion signal and is not impacted by the SSR.  The main benefit from squeezing comes from the reduction of noise away from the resonance, where the main contributions are the noise of the total amplification chain referred to the input of the \AMP, $N_{A}$, and the Johnson-Nyquist noise incident on and reflected off of the cavity, $N_{r}$,  which is modified by the cavity's reflection response.  Operation of the SSR creates correlations in the amplifier added noise and reflected noise, and appropriate choice of the amplifier pump phase yields a partial cancellation of the noise contributions, resulting in an improvement to the experimental sensitivity away from the cavity's resonance where these sources are dominant.

Because the resulting noise reduction is a function of \detune, the improvement to an axion search is quantified by comparing \Nsys to the expected power of an axion signal over a range of \detune's to capture improvements both on and off resonant with the cavity. The detected power is given by~\cite{Sikivie:1983ip_halotheory,Sikivie:1985yu_halotheory}
\begin{equation}
    \begin{split}
    \mbox{\axpwr}(\mbox{\detune}) = \left(\frac{2\pi \rho_{a} \hbar^{3} c^{3}}{ \mu_{0}} \right) & \left(\frac{g^{2}_{a\gamma\gamma}}{m^{2}_{a}} \right) \nu_{c} B^{2}_{0} V C_{nml}  \\
    &\times Q_{L} \frac{ \beta}{1+\beta} \frac{1}{1 +(2\mbox{\detune}/\Delta\nu_{c})^2}
        \label{eq:axion_power}
    \end{split}
\end{equation}
where $\rho_{a}$ is the local dark matter density \DarkMatterDen~\cite{Read2014}, $B_0$ is the amplitude of the magnetic field along the longitudinal axis of the cylindrical cavity, $V$ is unfilled volume of the cavity and $C_{nml}$ is the cavity form factor for the mode indexed by $nml$.  The cavity's linewidth, $\Delta \nu_{c}$, is given by the loaded quality factor as $Q_L=\nu_{c}/\Delta \nu_{c}$ which is reduced by a factor of $(1+\beta)$ from its unloaded value, $Q_{0}$, where the coupling of the antenna is parameterized by $\beta$.  As with $N_{c}$, this signal is defined as a function of \detune and is filtered by the cavity's transmission profile represented by the last term in parentheses.  While this power peaks exactly at \detune$=0$, there is still detectable power at \detune$>0$ proportional to the cavity's linewidth.  This results in each cavity tuning step having sensitivity to a range of axion masses.  This window, referred to as the measurement bandwidth, is defined as the region where the SNR,   
\begin{equation}
    \mbox{\SNR}(\mbox{\detune}) =  \frac{\mbox{\axpwr}(\mbox{\detune})}{h\nu \mbox{\Nsys}(\mbox{\detune})} \sqrt{\frac{\tau}{\Delta \nu}},
    \label{eq:vis_snr}
\end{equation}
exceeds a desired threshold. The scaling under the square root represents the improvement with the number of independent power measurement over a bandwidth $\Delta\nu$  taken over an integration period of $\tau$. 

A typical haloscope operates in a series of discrete tuning steps where subsequent spectra, whose bandwidths overlap, sum together to improve the total SNR.  Because of this, there is a benefit to increasing the bandwidth of each step by increasing the coupling $\beta$.  The resulting reduction of $Q_{L}$ decreases the experiment's sensitivity near the cavity's resonance, but can be compensated for by an increase in off resonant sensitivity if the SNR at these frequencies are above the desired threshold. For a given haloscope, the optimal overcoupling is defined as the $\beta$ which maximizes the scan rate
\begin{equation}
    R \propto \int_{-\infty}^{+\infty} \mbox{\SNR}^{2}(\mbox{\detune}) d\mbox{\detune}.
    \label{eq:scan_rate}
\end{equation}
Without squeezing the optimal overcoupling is found to be $\sim$2~\cite{Kenany2017design}, with operation beyond this point resulting in a reduction to the scan rate as the noise from $N_{r}$ quickly washes out signals for all frequencies. By making use of the SSR, the noise contribution from $N_{r}$ is reduced over a broad range of frequencies and improves the SNR at $\mbox{\detune}>0$.  This allows the cavity's coupling coefficient to be increased beyond the optimal value for an unsqueezed measurement, resulting in a larger measurement bandwidth for each tuning step and in turn gives an enhancement to the scan rate of the experiment. 

For an experiment which is able to deliver squeezing \emph{S}, defined as the reduction in the variance of the system output $S=\sigma^{2}_{\textrm{on}}/\sigma^{2}_{\textrm{off}}$, the optimal overcoupling is approximately $\beta$=2\emph{S}.  In the case of an ideal system, with no transmission loss, this can be used to arbitrarily increase the scan rate which approximately scales as $R \propto S$.  However, the deliverable squeezing in a realistic system is limited by transmission loss between the \SQZ and \AMP JPAs, which result in part of the squeezed state being replaced by unsqueezed vacuum.  For a system with transmission efficiency $\eta$ between the JPAs the theoretical limit to the delivered squeezing, assuming noiseless amplification, is given by
\begin{equation}
    S = \frac{\eta}{G_\text{\SQZ}}+1-\eta
    \label{eq:squeezing_limit}
\end{equation}
where $G_\text{\SQZ}$ is the one quadrature gain of the \SQZ. During HAYSTAC's first operation using the SSR in \PIIa a transmission efficiency of 0.63 was achieved with a \SQZ gain of \SI{13}{\dB}.  This resulted in a delivered squeezing of \SI{4}{\dB}, allowing for the scan rate to be roughly doubled relative to operation without squeezing~\cite{backes2021quantum}.

\subsection{Detector Design and Performance}
\label{sec:cavity}

\begin{figure}
    \centering 
    \includegraphics[width=0.5\textwidth]{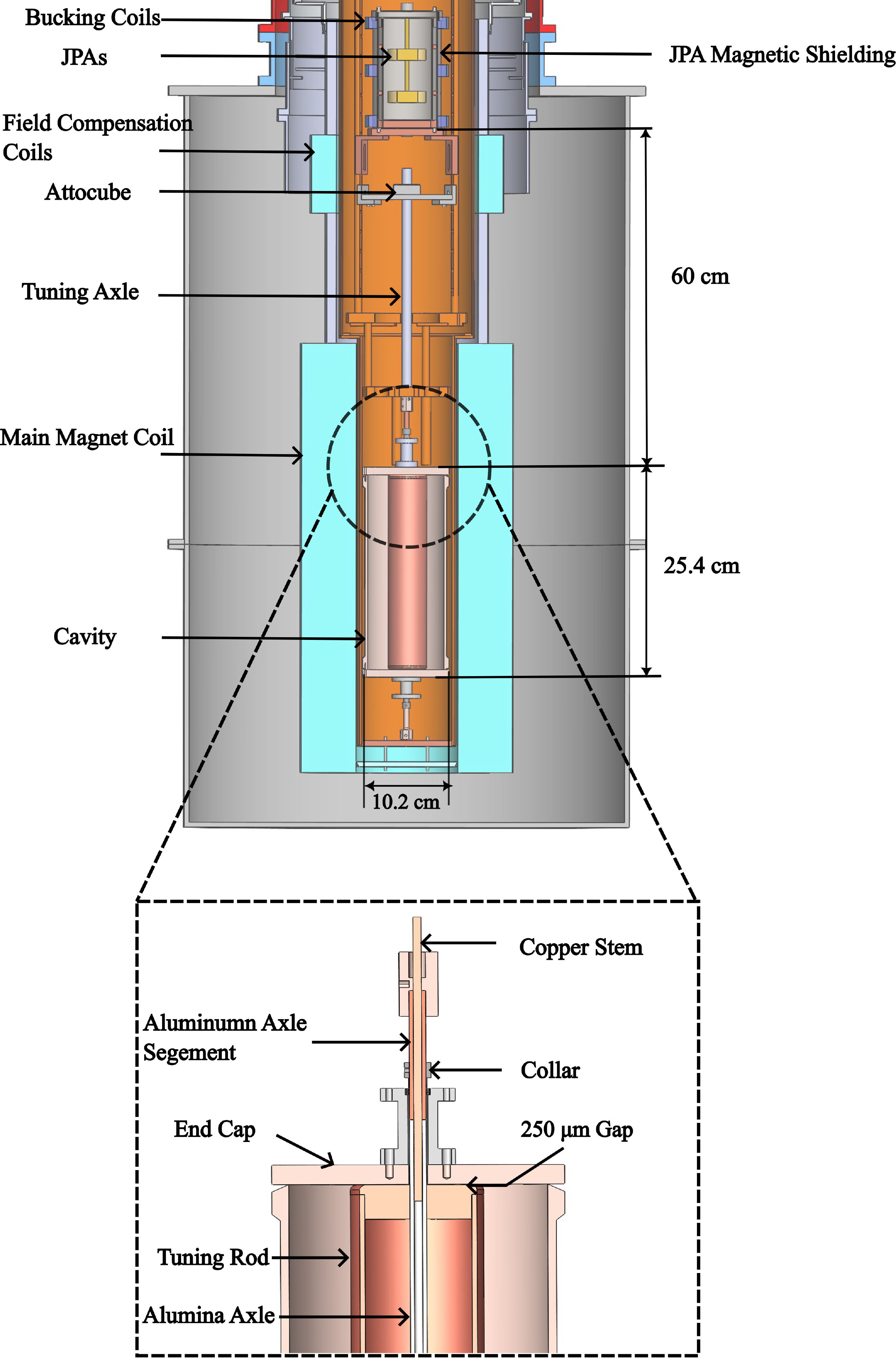}
      \caption{ CAD model of (top) the main cavity installed inside of the magnet's bore and (bottom) a detailed view of the top of the cavity highlighting the tuning rod design. The axle used to tune the cavity is segmented such that only the small section outside of either end-cap are conducting, with the sections inside of the cavity made of alumina.  The tuning rod is further cooled by inserting a copper stem through the hollow axle, from the top, bottom or both ends. Also shown is the JPA housing located \jpaheight above the main cavity. The magnet's \SI{70}{\kelvin} and \SI{4}{\kelvin} shields, and the fridge's  \SI{4}{\kelvin} and \SI{1}{\kelvin} shields are omitted for simplicity.}
    \label{fig:cavity_cad}
\end{figure}

A model of the experimental setup showing the cavity installed in the magnet's bore is shown in Fig.~\ref{fig:cavity_cad}.  The main cavity is composed of a cylinder of copper-plated stainless steel which is \cavlength long and has a \cavdiameter inner diameter. The resonant frequency of the cavity is tuned by rotating a \roddiameter diameter tuning rod, also made of copper-plated stainless steel, which pivots about an off-axis axle.  The rod takes up $\sim$25$\%$ of the cavity volume, leaving an unfilled volume of V$=$\cavaxionvolumeNoErr. Rotary motion of the axle is driven by an Attocube ANR240 precision piezoelectric motor \cite{attocube_2021} designed to operate at cryogenic temperatures.  The cavity mode of interest for the axion search is the TM$_{010}$, as this mode has the highest overlap with the external magentic field as represented by the form factor $C_{010}$. Varying the position of the rod by \SI{180}{\degree} allows for tuning over the range of \CavityFreqRange for the TM$_{010}$ mode. The experiment is operated inside an LD250 BlueFors dilution refrigerator \cite{bluefors_2020} which achieves a base temperature of \basetempNoErr at the mixing chamber plate.  The cavity sits inside the \MagBoreSize bore of a solenoidal magnet from Cryomagnetics Inc, used to generate an \magfield magnetic field within the cavity volume.  

To prevent degradation of the quality factor due to coaxial loss, the segment of the axle inside of the cavity and extending through the tuning rod is a \AxleDiameter diameter hollow alumina tube, which transitions to a short aluminum segment outside of the cavity on either end-cap as shown in Fig.~\ref{fig:cavity_cad} for the top end-cap.  The axle is fixed in place with collars on either side, which rest on the inner race of their respective bearings.   This results in a \rodgap gap between the rod and either end-cap to allow for smooth tuning of the rod.  Due to the low thermal conductivity of alumina at cryogenic temperatures \cite{alumina_2011,Drobizhev:2016qbx}, a copper stem with diameter \CuStemDiameter is inserted into the axle to improve the thermal link between the tuning rod and the cavity with an insertion depth chosen to maximize thermal contact area while simultaneously minimizing a potential decrease in quality factor.  This was found to be necessary during early operations of \PI which observed excess thermal noise from the cavity, consistent with a component of the cavity failing to reach base temperature~\cite{Kenany2017design,zhong2018results}. Including the stem reduced the effective noise temperature from \SI{600}{m\kelvin} to \SI{250}{m\kelvin} while still maintaining an unloaded quality factor of $Q_{0}\approx45\times10^{3}$ at cryogenic temperatures. \PIIa used separate stems inserted from both the top and bottom of the cavity, while \PIIb used a single stem from just the bottom. Both configurations achieved similar performance. While the cavity's effective noise temperature still appears elevated above the base temperature of \basetempNoErr, the exact cause is unknown with recent studies hinting at possible RF leakage through one of the cavity ports. Eliminating the remaining excess cavity noise is an ongoing effort in the HAYSTAC collaboration.

Located on the cavity's top end-cap are two antenna ports. The first is a weakly coupled port used for inputting microwave tones for calibration of the detector response while the second is a strongly coupled test port and used to read signals out of the cavity.  The depth of the strongly-coupled antenna is adjusted with a mechanically actuated stepper motor~\cite{appliedmotion_2022} which is attached to the antenna via a Kevlar line.  This allows for the antenna's coupling coefficient to be maintained near the optimal value of $\beta\sim7.1$ needed for the axion search to maximally benefit from the SSR~\cite{backes2021quantum}.  This is slightly larger than the estimate in Ref.~\cite{malnou2019squeezed} due to excess cavity noise as found during the noise calibrations described in Appendix~\ref{sec:system_noise}.

\begin{figure*}
    \centering 
    \includegraphics[width=\textwidth,trim={0 5.5cm 0 0},clip]{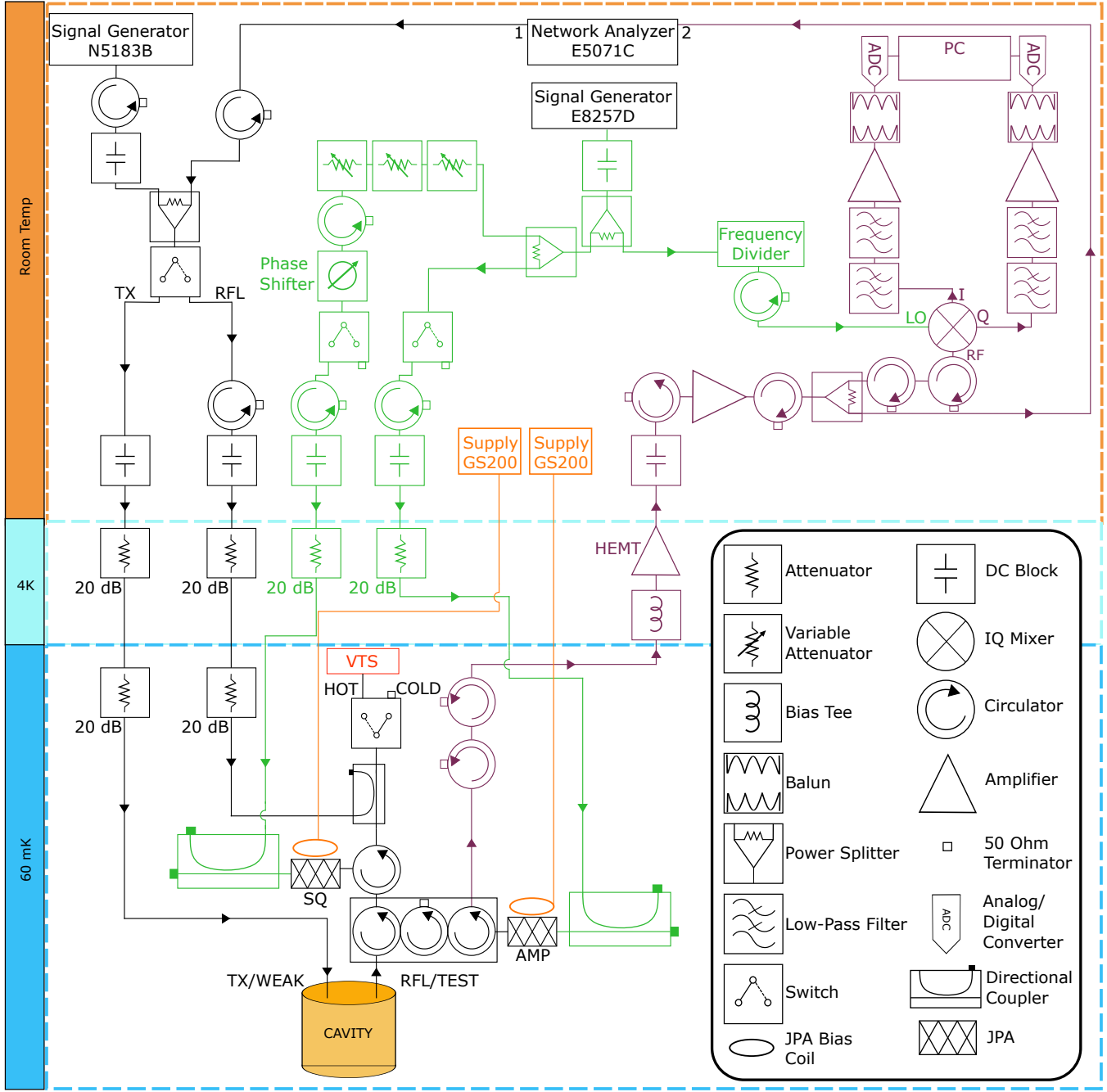}
      \caption{Schematic of HAYSTAC's electronics including both the IF and RF components. The electronics are divided by their temperature stage with the top stage held at room temperature, the middle at \SI{4}{\kelvin} and the bottom at \basetempNoErr.  Input lines into the system (black) are used to send signals for calibration and characterization of the system, while output lines (purple) are used to transfer signals from the output of the cavity to the DAQ to be recorded.  The lines providing the pump tone to the JPAs and mixer (green) and the lines for the JPA bias currents (orange) are also shown.}
    \label{fig:combined_circuit_v2}
\end{figure*}

To monitor the cavity's field, the two ports described above are coupled to the external receiver chain as shown in Fig.~\ref{fig:combined_circuit_v2}. This chain allows for power emitted by the cavity to be read out as well as for the injection of signals needed to calibrate the detector response. During operation the main input to the receiver chain comes from the thermal noise of a \SI{50}{\ohm} load which can be toggled by a switch between a cold load held at the base temperature of the mixing chamber and a hot load held at an elevated temperature by a Variable Temperature Stage (VTS) which operates at temperatures between \SI{290}{m\kelvin} and \SI{1}{\kelvin}. The temperature of the VTS is measured by an MFFT-1 Magnicon temperature sensor~\cite{magnicon_2021}. During \PIIa(\PIIb) this temperature was maintained at \PIIaVTS(\PIIbVTS) for use in calibration measurements described in Appendix~\ref{sec:system_noise}. The VTS was incorporated prior to the start of \PII allowing a finer temperature control of the hot load, previously limited to the temperature of the still plate, $\sim$\SI{775}{m\kelvin}. 

Signals in the receiver chain are routed using 0.085'' coaxial cables with the outer conductors of the cables thermalized at each temperature stage with gold-plated copper clamps which also serve to block stray light coming from higher temperature stages.   The cable segments above the \SI{4}{\kelvin} stage use stainless steel cables while segments below this use superconducting NbTi/NbTi cables.  The extremely small losses for the superconducting coax was experimentally shown to be independent of applied fields up to \SI{7}{\tesla}, due to shielding of the inner surfaces from the applied field by the superconducting jacket. This low-loss behavior can be expected up to H$_{c2}$=\SI{14}{ T}. In addition, DC electrical signals, used to tune the JPAs and power the HEMT, are sent into the fridge through API 56F715-005 EMI filters mounted on DB15 feedthroughs.  Inside the fridge these DC signals are routed to the still plate using 28 AWG phosphor-bronze wire before transitioning to superconducting wire.

Upon exiting the fridge, the output signal from the cavity is split into two paths with one path leading directly to the return port of an Agilent E5071C vector network analyzer~(VNA)~\cite{agilent_vna}, while the other leads to a PC with a 14-bit GaGe Oscar CSE4344 PCIe  digitizer~\cite{noauthor_14-bit_nodate} used to sample the cavity field.  Prior to reaching the digitizer the signal is sent to an IQ-mixer (Marki Microwave IQ0307LXP~\cite{marki_iqmixer}) where it is split into the I/Q quadratures and down-converted into an intermediate frequency (IF) band centered at the cavity's resonant frequency or equivalently half the JPA pump frequency.  To precisely match these frequencies, the same signal generator is used as both the pump for the JPAs as well as the local oscillator~(LO) for the mixer. A frequency divider before the mixer is used to convert the tone down to the cavity frequency.  Finally, before reaching the digitizer the signal from both quadratures pass through low-pass filters which cut off the IF window at \SI{1.9}{\MHz} in order to remove high frequency noise outside the cavity's usable bandwidth.  All connections between these IF components are made with double-shielded BNC cables and the signal generators, VNA, and digitizer are all clocked to the same \SI{10}{\MHz} rubidium reference source (SRS FS725m~\cite{srs_rubidium_ref}).

\begin{figure}[ht]
%analysis/PhaseIIab_paper/plot_tx_rfl.ipynb
    \centering 
    \includegraphics[width=\columnwidth]{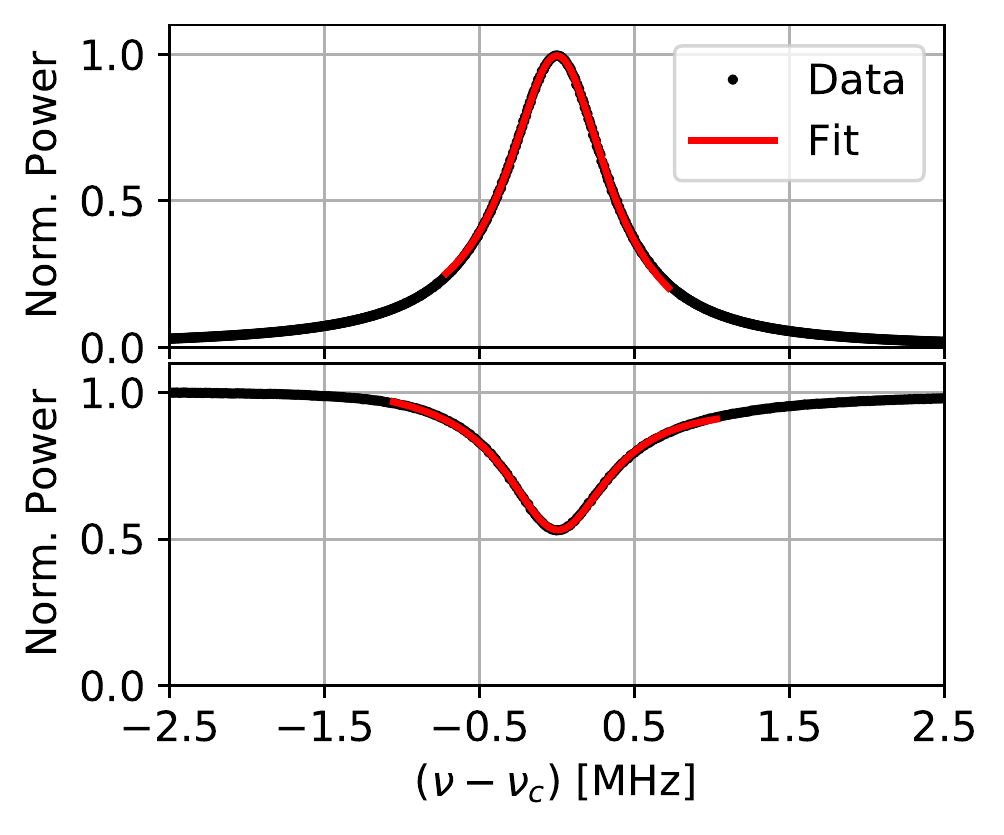}
      \caption{Example transmission (top) and reflection (bottom) measurements (black) taken with the VNA in \PIIb along with the fit (red) to Eq.~\ref{eq:lorenz} used to extract the relevant cavity parameters.  The profile for both is normalized, with the transmission profile normalized to have unit peak power and the reflection normalized to unit power off resonance.}
    \label{fig:tx_rfl_fits}
\end{figure}

\begin{figure}[ht]
    \centering 
    \includegraphics[width=\columnwidth]{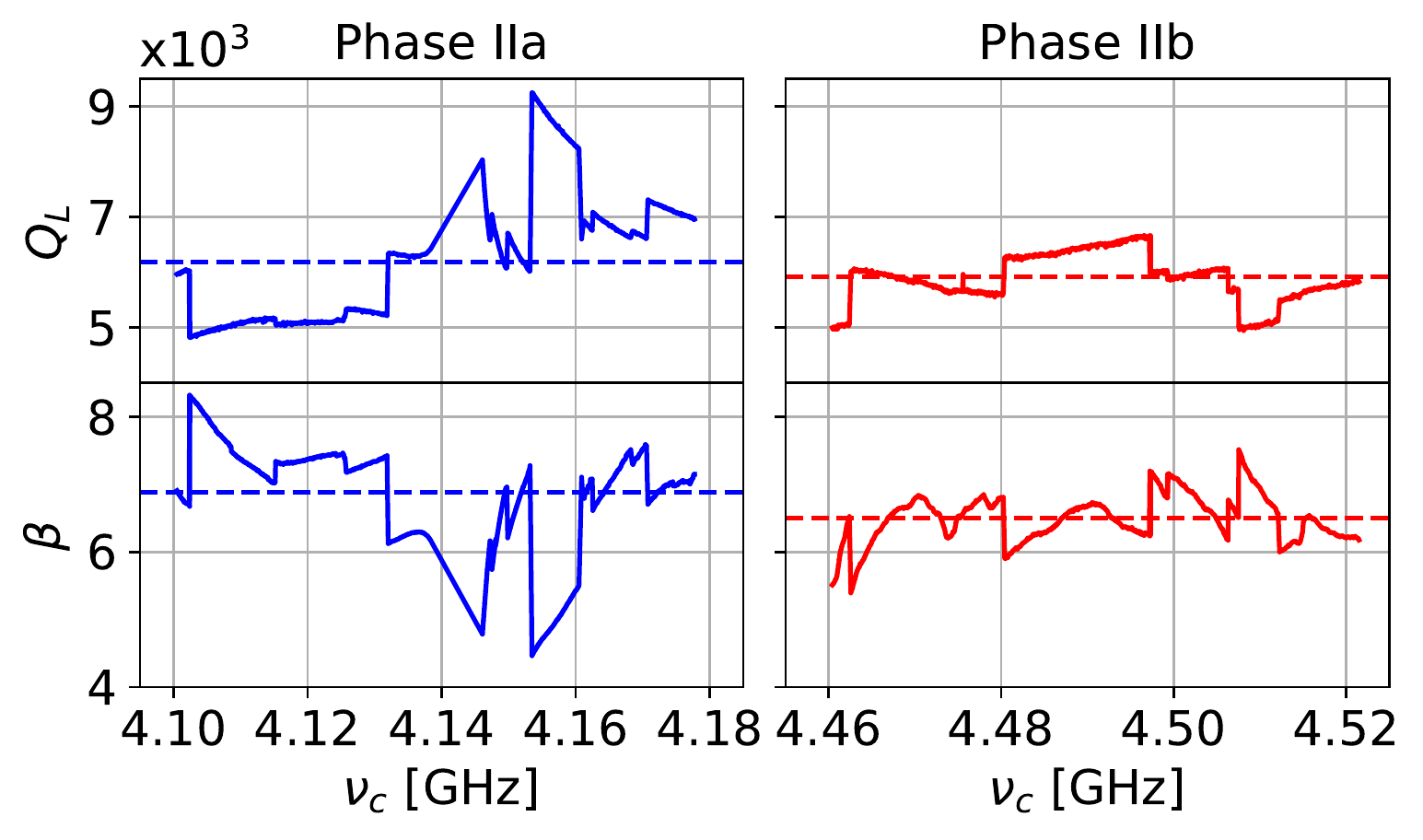}
      \caption{Measured values of $Q_{L}$ and $\beta$ as a function of $\nu_{c}$ for every run taken during the original scans in both \PIIa (left) and \PIIb (right) with the average value over each phase shown as a dashed line.  The discontinuous jumps in both values correspond to adjustments of the antenna position to maintain $\beta$ near the optimal value.}
    \label{fig:q_beta_trend}
\end{figure}

The cavity's performance is characterized with a VNA, which injects stimuli into either of the two ports described above.  Using a pair of VNA measurements, the cavity's TM$_{010}$ mode frequency $\nu_{c}$, coupling coefficient $\beta$ and loaded quality factor $Q_{L}$=$Q_{0}/(1+\beta)$ can be extracted.  In both cases the observed VNA response is fit to the expected functional form of the power
\begin{equation}
    P(\nu) = \frac{P_{0}}{\left(2Q_{L}\left(\nu/\nu_{c} - 1\right) \right)^{2} + 1} + \left( A\nu + B \right), 
    \label{eq:lorenz}
\end{equation}
where the first term is a Lorentzian with peak amplitude $P_{0}$ and the second term in parentheses represents a linear baseline.  The first is a transmission measurements~("tx"), taken by sending a stimulus to the weak port and reading it out through the test port via the cavity and is used to extract both $Q_{L}$ and $\nu_{c}$.  Following this, a reflection measurement~("rfl") is taken by routing the stimulus signal to the strongly coupled port where it is reflected off of the cavity and back to the VNA.  This is used to extract $\beta$ as 
\begin{equation}
    \beta = \frac{1 + \sqrt{P(\nu_{c})/(A\nu_{c} + B)}}{1 - \sqrt{P(\nu_{c})/(A\nu_{c} + B)}},
\end{equation}
when $\beta>1$~\cite{pozar_microwave_2011}.  Example measurements taken during \PIIb, along with their corresponding fits, are shown in Fig.~\ref{fig:tx_rfl_fits} and the resulting values of $Q_{L}$ and $\beta$ at each frequency scanned in \PII are shown in Fig.~\ref{fig:q_beta_trend}.  Their values vary as the cavity frequency changes, with the largest variations coming from discontinuous jumps corresponding to adjustments of the antenna position to maintain the coupling coefficient near the value which optimizes the scan rate.  The average values over \PIIa(\PIIb) are $Q_{L}=6180(5915)$ and $\beta=6.9(6.5)$.         

During \PIIa operation the TM$_{010}$ mode exhibited residual drift, taking $\sim$\SI{10}{\minute} to settle to its new resonance each time the rod position was adjusted.  To avoid degrading the detector's sensitivity, a \SI{15}{\minute} delay was imposed after each tuning step during which no data was taken, slightly decreasing the detector livetime.  This effect worsened over time such that during the acquisition of rescan data the tuning between each target candidate became inconsistent and typically required manual intervention.  While neither of these issues had a major impact on the results of \PIIa, the cavity was reassembled and the tuning rod realigned to increase its stability prior to the start of \PIIb. As a result, the cavity tuning was more reliable in \PIIb  and the observed mode drift decreased substantially.  This allowed the wait time to be reduced from \SI{15} minutes to \SI{5} minutes, chosen conservatively to ensure that heat produced from tuning is dissipated.

\subsection{Design and Performance of the Squeezed State Receiver}
\label{sec:JPAs}

\begin{figure}
    \centering 
    \includegraphics[width=\columnwidth]{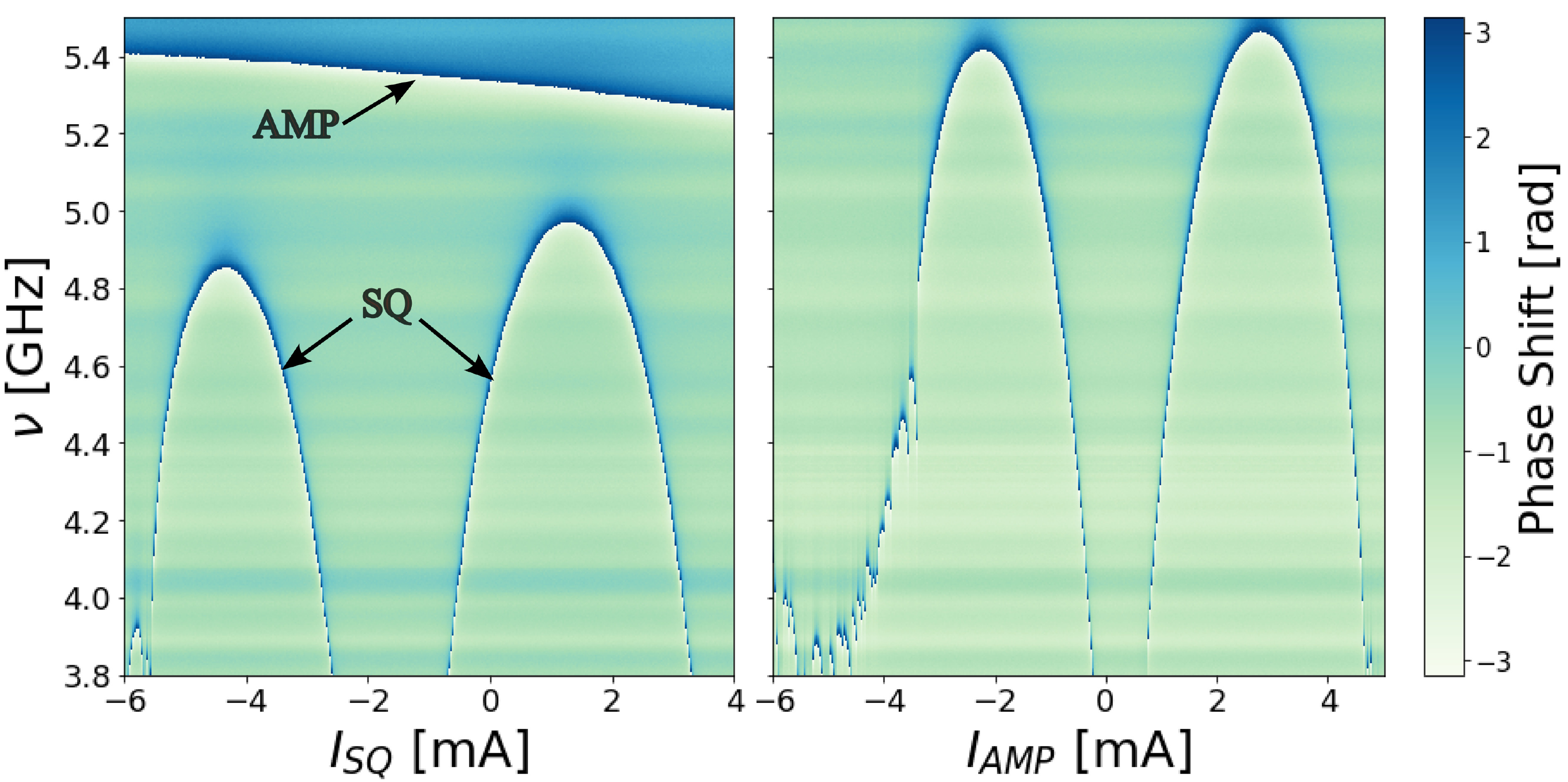}
      \caption{Example flux tuning curves of the \SQZ JPA~(left) and the \AMP JPA~(right) produced from measuring the phase response as a function of frequency $\nu$ with the VNA while varying the current $I_{JPA}$ in one of the JPA's bias coils. The resonance of the JPA at a given $I_{JPA}$ is the frequency at which the phase undergoes a $\pi$ phase shift.  In the tuning curve for the \SQZ, the \AMP's curve is also present at $\sim$\SI{5.3}{\GHz} but stays roughly constant due to reverse biasing the \AMP's current. The observed residual variation is due to an initially imperfect compensation and was corrected before axion data was taken.  The \SQZ curve is also present in the \AMP's curve but is below the lowest frequency plotted here. Additionally, the non-smooth regions in the curves are likely due to non-uniformity of the magnetic flux penetrating each JPA.}
    \label{fig:jpa_tuning}
\end{figure}

The SSR used in this work follows a similar design to that used in \cite{malnou2019squeezed} and is composed of two JPAs operated in phase-sensitive mode which are coupled to the main cavity via the test port as shown in Fig.~\ref{fig:combined_circuit_v2}. Each JPA acts as a nonlinear \emph{LC} resonator, where the nonlinear inductance is achieved by an array of superconducting quantum interference devices (SQUIDs). The JPAs used in this work are flux-pumped JPAs which use an on-chip bias line to modulate the magnetic flux threading the SQUID loop at twice the resonance frequency~\cite{yamamoto2008flux}. The resonance of each JPA is then set by a static flux provided by a pair of superconducting bias coils on which each JPA is mounted.  The flux through each bias coil is set separately by supplying a current from one of two identical current sources~(Yokogawa GS200~\cite{yokogawa}),  allowing the resonance frequency of each JPA to be adjusted semi-independently. To determine the operating range for each JPA, a flux tuning curve that correlates the JPA's resonant frequency with coil current is created by sending stimulus signals from the VNA to the JPA. By measuring the phase response of the transmitted signal, a tuning curve is extracted by finding the point at which the phase changes by $\pi$. Example tuning curves for the two JPAs used in \PIIb are shown in Fig.~\ref{fig:jpa_tuning}. Comparing the two curves, a common region of gain is observed between 4.2--\SI{4.8}{\GHz}. While an ideal JPA would be periodic in its response, region of non-ideal behavior are observed in Fig.~\ref{fig:jpa_tuning}.  Although the exact cause of these regions is unknown, it is likely the result of flux trapped in the chip's superconducting circuitry or by other sources of non-uniformity in the magnetic flux penetrating each JPA.  To achieve stable gain, these regions are avoided when operating the JPAs.  For the JPAs used in \PIIa this operating range only extended to $\sim$\SI{4.2}{\GHz} by design, to extend the search to higher frequencies a new set of JPAs were obtained and installed prior to the start of \PIIb. 

Once the flux bias needed to obtain the desired resonant frequency is found, amplification is achieved by pumping each JPA at twice this frequency with a single low noise signal generator~(PSG E8257D~\cite{psg_pump}) shared by both JPAs as shown in Fig.~\ref{fig:combined_circuit_v2}. The absolute amplification gain is controlled by the power of the pump tone sent to each JPA. An automated procedure, which scans over a range of coil currents and pump powers to find the optimal parameters for the desired gain, is used to individually bias each JPA during operation of the experiment. 

Complications to this approach not present in the test setup from Ref.~\cite{malnou2019squeezed}, which used an injected signal and did not require a magnetic field, arise from the proximity of the JPAs to the large magnetic field required to convert axions into a detectable signal.  To minimize the impact, the JPAs are held \jpaheight above the cavity as shown in Fig.~\ref{fig:cavity_cad}.  Here the magnetic field is reduced to $\sim$\SI{50}{G} by a pair of NbTi field compensation coils, oppositely oriented and in series with the magnet. Further shielding from the remaining stray field is achieved in a multi-staged shielding configuration. First, a set of three persistent superconducting NbTi coils are mounted around the amplifier canister to counter changes to the magnetic flux. Inside of these coils, the JPAs are housed in a four-layered shielding can. From inside out, this consists of a layer of Amumetal \SI{4}{K}, aluminum, Amumetal \SI{4}{K} and niobium. In this way the Amumetal provides shielding as the JPAs are cooled to cryogenic temperatures, at which point the aluminum and niobium layers become superconducting and provide additional shielding. Together this reduced stray flux to $<<$1 flux quantum over the area of the JPAs' SQUID loops, negligible to the operation of the JPAs as confirmed by measurements of the tuning curves with and without the field which show no measurable change. Due to limitations of space inside the fridge, the two JPAs are housed inside of the same shielding can and are located $\sim$\SI{10}{\cm} apart with no shielding between them. As a result of their close proximity, changing the current of one coil affects the flux in both JPAs and biasing either JPA independent of the other is not possible. To decouple them, the residual flux induced in the secondary JPA (the JPA that is not actively being tuned) is empirically measured as a fraction $\epsilon$ of the change of the primary JPA (the JPA that is actively being tuned) current.  This allows for the secondary JPA's frequency to be held constant by applying a compensating current $\epsilon \Delta I$ each time the primary JPA's current is changed. This phenomenon is visible in Fig.~\ref{fig:jpa_tuning} (left), where the \AMP's curve at $\sim$\SI{5.3}{\GHz} is slowly tuned as the bias for the \SQZ is changed due to an initially imperfect compensation.

\begin{figure}
    \centering 
    \includegraphics[trim={1.3cm 0cm 4cm 0cm},clip, width=\columnwidth]{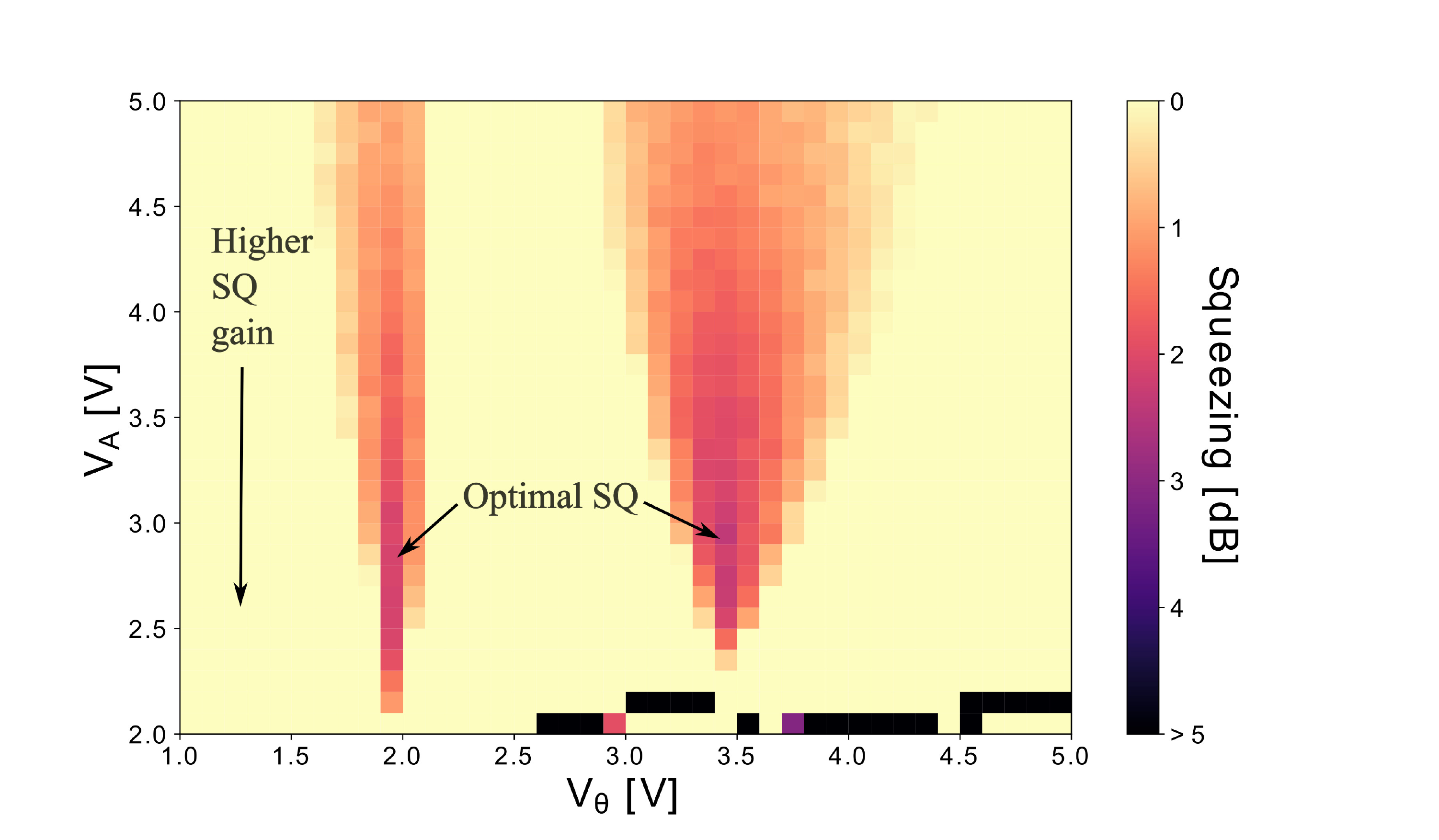}
      \caption{Example plot of delivered squeezing mapped over both the voltage of the variable phase shifter $V_{\theta}$ and variable attenuator $V_{A}$.  This mapping allows for the identification of the the $V_{\theta}$ and $V_{A}$ which maximize squeezing, where the optimal $V_{\theta}$ corresponds to a $\pi/2 + n\pi$ phase difference between the \SQZ and \AMP and the optimal $V_{A}$ corresponds to the largest \SQZ gain before saturation.  Additionally, beyond a certain gain this fully saturates the chain causing regions of fake squeezing seen here as the disconnected points at $V_{A}<2.1$\si{V}.}
    \label{fig:sq_colormap}
\end{figure}

As described in Sec~\ref{sec:ssr_det_principle} vacuum fluctuations in the experiment are squeezed by operating the two JPAs at the same resonant frequency, with a $\pi/2$ phase shift between the two pump tones such that their amplified/squeezed quadratures are perpendicular. This is achieved by using a single signal generator to generate the tone used by both JPAs, with a variable phase shifter located before the squeezer used to apply a phase shift proportional to the voltage $V_{\theta}$. In addition, a variable attenuator is added to the tone sent to the \SQZ JPA to allow for its gain to be set independently from the \AMP JPA by varying the applied voltage $V_{A}$.  This is required as the optimal \SQZ gain is typically $\sim$\SI{15}{\dB} lower than that of the \AMP to avoid potential saturation~\cite{malnou2019squeezed,Malnou:2017udw}.  As a result, operating the SSR requires finding the pair of $V_{A}$ and $V_{\theta}$ for which the squeezing is optimal.  This is done by scanning over a range of $V_A$ and $V_{\theta}$ and collecting short (\SI{10}{ms}) time traces with data acquisition software, first with the SQ on and then with it off. Squeezing is then quantified by the ratio of the variances of the two time traces ($S=\sigma_\text{on}^2/\sigma_\text{off}^2$). This results in a map of squeezing $S$ over a range of $V_A$ and $V_{\theta}$ as shown in Fig.~\ref{fig:sq_colormap}.  This  allows for the set of parameters which result in the optimal squeezing to be found.  The example shown here is from a coarse scan used to approximately find the location of the optimal parameters and is used to initialize the finer scan performed during acquisition of axion data. This finer scan is done in a small region of the full parameter space using an automated tuning algorithm that searches in the vicinity of the initial guess in decreasing voltage step sizes in order to narrow down both $V_A$ and $V_{\theta}$ until the optimal value of $S$ is reached.  Because the tuning step size of the experiment ($\sim$\SI{100}{\kHz}) is small relative to the change in these variables with frequency, the most recent set of optimal parameters are used to initialize the algorithm each time the cavity is tuned allowing for efficient tuning of the SSR.   

\begin{figure}[t]
    \centering 
    \includegraphics[width=\columnwidth]{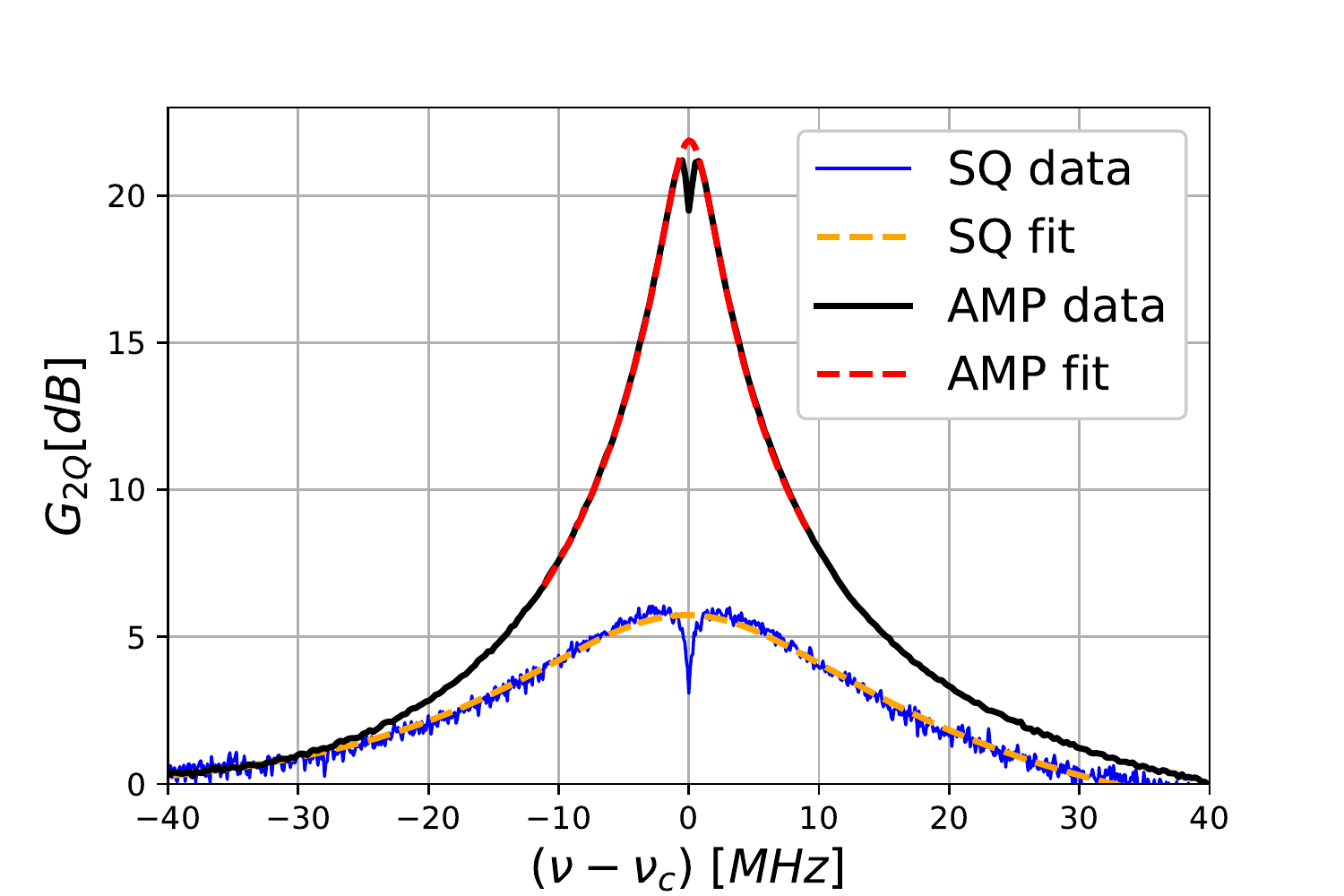}
      \caption{Example 2Q gain profiles of the \AMP and the \SQZ as measured by the VNA (solid lines) and their fits to a Lorentzian (dashed lines). Because the data is taken through rfl and the JPA is centered on the cavity's resonant frequency, the cavity's reflection profile appears as a dip at \SI{0}{\MHz}.  When fitting the profile the $\sim$\SI{2}{\MHz} region where this effect is most prominent is ignored so as not to distort the fit.}
    \label{fig:jpa_amp_prof}
\end{figure}

Following this optimization, the gain along the amplified quadrature as well as its frequency dependence are measured.  In order to measure the gain in just one quadrature~(1Q) a calibration tone is generated by a signal generator~(N5183B MXG~\cite{mxg_pump}) and injected into the system through the "rfl" path at a detuning of \SI{200}{\kHz} relative to the JPA's resonance. The ratio of the observed power with and without the JPA of interest being pumped, as measured by the digitizer, gives a measurement of the gain along the amplified quadrature at this one detuning. Because the 1Q gain measurement is time consuming, the full frequency dependence of the gain at all detunings is mapped out with use of the VNA which sweeps a stimulus signal through many frequencies along the same path.  This results in a measurement of the two-quadrature~(2Q) gain profile, an example of which is shown from \PIIb for both the \AMP and \SQZ in Fig.~\ref{fig:jpa_amp_prof}.  In \PII the target 2Q gains were $\sim$\SI{22}{dB} for the \AMP and $\sim$\SI{5}{dB} for the squeezer. The \AMP gain is chosen such that it is large enough to overwhelm the added noise of the high-electron-mobility transistor~(HEMT) which follows the AMP in the receiver chain. 
 The target gain of the \SQZ JPA is kept low enough to prevent saturation of the \AMP JPA. 

For the SSR used in HAYSTAC the delivered squeezing off-resonant from the cavity is measured prior to the start of each run to be \SI{3.7}{dB} in \PIIb and \SI{4}{dB} in \PIIa.  This level of squeezing is consistent with the expectation given by Eq.~\ref{eq:squeezing_limit} given the measured transmission efficiency, $\eta$, of \PIIbEta in \PIIb and \PIIaEta \PIIa.  This efficiency is largely dominated by losses in the custom triple-junction circulator~(QuinStar QCE-070100CM30~\cite{noauthor_quinstar}) located between the \SQZ and \AMP JPA in Fig.~\ref{fig:combined_circuit_v2} which is used to isolate the JPAs from unwanted noise reflecting backwards through the system.  The difference between the transmission in the two phases is likely due to the frequency dependence of the circulator's loss. Assuming this level of squeezing is maintained over the full frequency range, the scan rate is expected to increase by a factor of 1.9 in \PIIb and 2.0 in \PIIa relative to an equivalent measurement taken in the same configuration but without squeezing.  To precisely calibrate the noise of the system, the performance of the SSR and its delivered squeezing are monitored periodically during data taking as described in Appendix~\ref{sec:system_noise}.

\subsection{Data Acquisition}
\label{sec:DAQ}

Once the signal reaches the digitizer, the voltage of both IF channels are sampled  at a rate $f_{s}$ and the sampled data is transferred to the RAM of the data acquisition (DAQ) PC.  The rate at which the data is sampled is chosen such that the Nyquist frequency is comfortably above both the \SI{1.9}{\MHz} cutoff of the IF filters as well as  the cavity's sensitive bandwidth roughly given by 2$\Delta\nu_{c}\sim$\SI{1400}{\kHz}.  In \PIIa a sampling rate of \SI{10}{\MHz} was used but was lowered to \SI{5}{\MHz} in \PIIb in an effort to reduce the total amount of data that is transferred to RAM.

Before saving the data to disk it is first processed \emph{in situ}.  The reason for this it that at these sampling rates the raw data from a single a single \SI{1}{hour} observation would require roughly $\sim$\SI{60}{\giga\byte} of disk space.  Rather then save all of this data to disk, the data is processed such that only a fraction of the original data is stored.  The processing begins by sampling the voltage fluctuations $V_{I,Q}(t)$ of both quadratures output by the mixer in \SI{5}{second} segments.  Because these quadratures are those coming from the mixer, they are not necessarily aligned to the amplified and squeezed quadratures of the SSR and must be rotated by an angle $\theta_\text{AS}$ by the linear combinations
    \begin{equation}
        V_\text{AMP}\left(t\right) = V_I\left(t\right)\sin\theta_\text{AS} + V_Q\left(t\right)\cos\theta_\text{AS}
    \end{equation}
for the amplified quadrature and 
    \begin{equation}
        V_\text{SQZ}\left(t\right) = V_I\left(t\right)\cos\theta_\text{AS} - V_Q\left(t\right)\sin\theta_\text{AS}
    \end{equation}
for the squeezed quadrature.  This angle is empirically determined at each tuning step as the angle which maximizes the variance of $V_\text{AMP}$, indicating alignment with the amplified quadrature.

The resulting voltage fluctuations of the amplified and squeezed quadratures are further divided into sub-segments consisting of $N_{s}$ samples and the power spectrum of each is computed from the FFT.  For \PIIa a sub-segment size of $10^5$ was chosen but this was reduced to $2^{15}$ in \PIIb.  In both cases this resulted in a frequency resolution  ($\Delta \nu_{b}=f_{s}/N_{s}$) which was smaller then the \AxionLineWidth expected axion linewidth $\Delta \nu_{a}$, but the reduction in \PIIb sped up the FFT computation.  The power spectrum of each sub-segment is then combined to produce an average power spectrum for the entire \SI{5}{\second} period.  These steps are repeated until the total observation time reaches the desired level, chosen to be \SI{1}{\hour} for both \PIIa and \PIIb, with the average power spectrum from each \SI{5}{\second} period being further combined to produced a single power spectrum representing the average over the full observation time.   

As a result of this processing, the data storage requirement is reduced by a factor of $\sim 10^{5}$ but requires usage of PC resources also needed by the digitizer.  This introduces deadtime during which voltage fluctuations from the cavity are not being recorded and reduces the effective scan rate of the experiment.  The deadtime from the \emph{in situ} processing, defined as the fraction of the total observation time spent not recording voltage fluctuations from the cavity, was \SI{40}{\percent} in \PIIa. In addition, time spent tuning the cavity and biasing the JPAs resulted in an additional \SI{12}{\percent} deadtime in \PIIa resulting in a total deadtime of \PIIaDeadPercent.

Several improvements were implemented for \PIIb resulting in a reduction of the total deadtime by a factor of $~\sim$1.6.  Two improvements were made to speed up the data acquisition  The first was to parallelize the data transfer with use of the two separate buffers on the digitizer card.  By swapping between these buffers, data can be transferred to disk while it is simultaneously being sampled, resulting in a reduction in the deadtime. The second improvement was to decrease the 
time needed to compute the power spectrum by both parallelizing the FFT algorithm with use of an NVIDIA Titan V GPU~\cite{NVIDIATitanV} while also reducing the sampling frequency as described above. While this did reduce both the frequency resolution and the width of the FFT, both still comfortably fit the requirements for an axion search previously described. In total the deadtime associated with the data acquisition was reduced by a factor of $\sim10$. In addition to improvements to the \emph{in situ} processing, the reassembly of the cavity prior to the start of \PIIb described in Sec.~\ref{sec:cavity} resulted in a significant reduction to the mode drift of the cavity.  As a result, the time needed for the cavity to settle after each tuning step was reduced by a factor of $3$.  All together, the improvements to the tuning and DAQ allowed for the total deadtime to be reduced to \PIIbDeadPercent resulting in an effective scan rate enhancement of $\sim$1.6 in \PIIb.

\subsection{Data Processing and Analysis}
\label{sec:analysis}

\begin{figure}[ht]
    \centering 
    \includegraphics[width=\columnwidth]{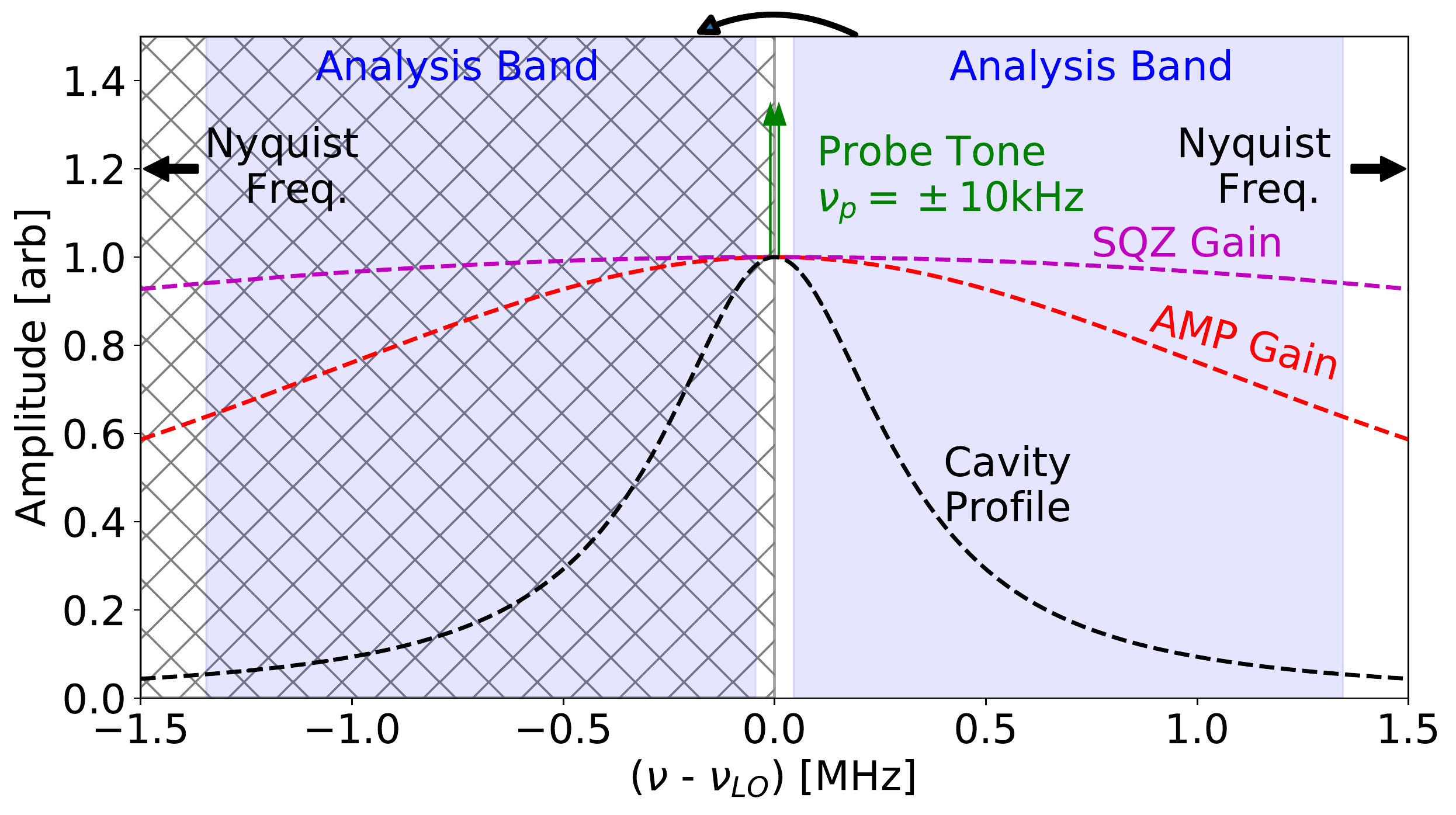}
      \caption{Diagram of the relevant frequencies for HAYSTAC's data taking and analysis, showing the relative shape of the cavity (black), \AMP(red) and \SQZ(purple) profiles as a function of detuning and normalized to the same peak height.  The measurement is taken in a homodyne configuration, with all the profiles centered at \SI{0}{\MHz}.  As a result, each bin contains power from both positive and negative detunings (Appendix~\ref{sec:system_noise}) and the double-sided profile is recovered by mirroring (grey cross-hatches) the spectrum around \SI{0}{\MHz}. Also shown is the analysis band (blue shaded region) and the probe tone (green) at \SI{10}{\kHz} used to monitor gain fluctuations over the course of each measurement.}
    \label{fig:analysis_band}
\end{figure}

Following acquisition of the average power spectrum at each tuning step data is processed and analyzed in order to search for potential candidates, identified as power excesses in the spectrum above a chosen threshold.  The following section gives a brief outline of this procedure.  A more detailed procedure is found in~\cite{brubaker2017analysis}, with the main difference arising from the homodyne configuration which centers the cavity around the JPA's resonant frequency.

The first step of this procedure is to combine the average power spectrum observed at each tuning step.  While each of these spectra individually have sensitivity to axion candidates, the benefit of having a larger bandwidth comes from the overlap between subsequent spectra which allows for the sensitivity at a given frequency to continue increasing even as it is detuned from the cavity.  In order to take advantage of this benefit, the power observed at each tuning step must be combined to form a single power spectrum which optimizes the SNR at each frequency. 

Prior to their addition, each spectrum must be filtered to remove spectral features that are not indicative of the axion's presence. This procedure starts by removing 
such structure shared between all power spectra in the IF. To facilitate this, spectra from different tuning steps are first aligned along the IF band and summed to produce an average IF response.  Because the IF band is by definition centered around the resonant frequency of the cavity and JPA, this response largely captures the spectral features of the cavity noise and the JPA gain. Signals in the RF, like those resulting from an axion, would be washed out as they would move in the IF band relative to the mixing frequency and effectively average away.  By applying a Savitzky-Golay (SG) filter~\cite{sg_filter} (width = 50kHz, order = 10) to the IF response the large scale spectral shape can be extracted and is subsequently removed from each spectrum by dividing out the filter's output. Additionally, contaminated IF frequencies can be identified  by comparing the average response before and after the filter.  Because noise from RF signals would average away in this procedure, any spikes in the average response which deviate from the filtered average must be due to IF noise which would add coherently.  In addition, each spectrum is trucated into the analysis band defined by a lower bound at \SI{45}{\kHz} to remove low frequency 1/f noise and an upper bound at \SI{1345}{\kHz}$\sim$2$\Delta\nu_{c}$ which covers the majority of the cavity's measurement bandwidth while staying below both the Nyquist and low pass cutoff frequencies.

\begin{figure*}
    \centering 
    \includegraphics[width=\textwidth]{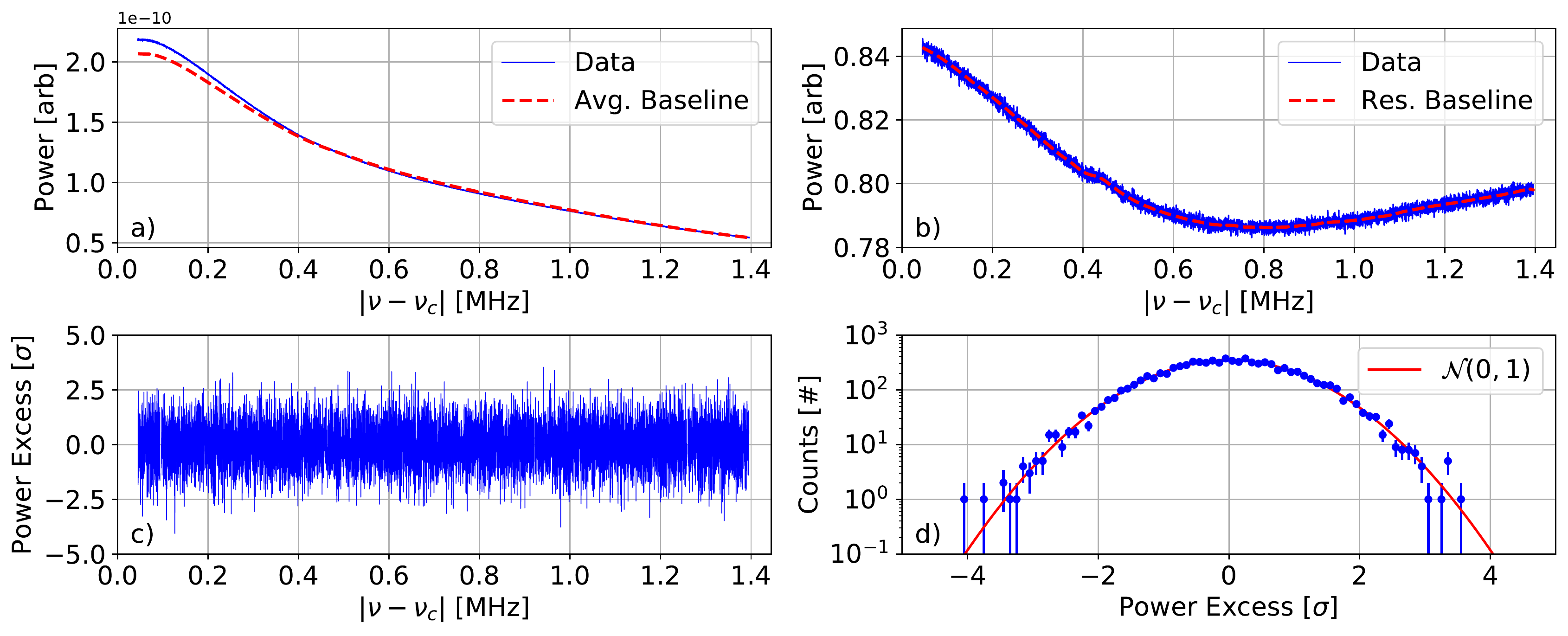}
      \caption{Example steps of the analysis procedure described in the text for a single spectrum taken during \PIIb.  (a) The raw data (solid blue) for a single tuning step is compared to the average baseline (dashed red) calculated from the full data set. The shape of both follow from the cavity's transmission profile with differences between the two arising from variations of experimental properties between each tuning step. (b) The same spectrum after dividing out the average baseline (solid blue) and removing IF contaminated bins along with its calculated residual baseline (dashed red). (c) The same spectrum after removing of the residual baseline and normalized such that each bin is approximately sampled from a normal distribution with $\mu=0$ and $\sigma=1$. (d) The distribution of the observed power in each frequency bin (blue points) from (c) showing good agreement with $\mathcal{N}(0,1)$ (solid red).}
    \label{fig:analysis_steps}
\end{figure*}

While this effectively removes a majority of the non-axionic spectral features, variations of the frequency-dependent profiles of both the JPA and cavity result in residual structure unique to each spectrum remaining, as seen in Fig.~\ref{fig:analysis_steps}.  Because both the cavity and JPA have bandwidths which are $>>\Delta\nu_{a}$, this residual structure can be effectively removed with minimal effect to the axion signal by applying a lower-order SG filter (width=\SI{50}{\kHz}, order=4) to each spectrum and dividing out the the filtered response from the original spectrum.  With these parameters the filter has a \SI{3}{dB} point around $\sim$\SI{60}{\kHz}, which is far enough above the $\sim$\SI{5}{\kHz} axion linewidth to avoid significant attenuation of its signal.  The resulting attenuation to an axion signal is found to be 0.91 using a simulation similar to that described in Ref.~\cite{brubaker2017analysis}. Following this, the resulting spectra consist of Gaussian-distributed noise with a mean of 0, as shown in Fig.~\ref{fig:analysis_steps}. 

The final step before summing the spectra from each tuning step is to weight the contribution of each IF bin of individual spectra by their observed SNR, so as to produce a sum spectrum which optimizes the total SNR. These weights are calculated separately for each spectrum with parameters extracted from the noise and signal characterizations described in Appendix~\ref{sec:uncer}.  Following this scaling, the spectra are aligned in the RF and summed to produce a single combined spectrum.  Because the data comes from a homodyne measurement in which the JPA's pump tone is also centered on the cavity's resonance, each frequency bin in the output spectrum contains contributions from both positive and negative detunings from cavity resonance. The double sided profile, shown in Fig.~\ref{fig:analysis_band}, is recoverd by mirroring each spectrum and its associated SNR around \SI{0}{\kHz}.  This allows for observations from both the positive and negative detuned frequencies to contribute to the combined spectrum with the appropriate weighting.

In the absence of any RF tones, this procedure results in a final combined spectrum which is composed of normally distributed noise fluctuations, with any axion-like signal appearing as a power excess with shape given by the expected virialized axion lineshape in the terrestrial lab frame \cite{lineshape1990}.  Because the resulting spectrum still has frequency resolution given by the DAQ and is intentionally set much smaller then the axion linewidth, an axion induced signal would appear divided across multiple bins.  To account for this, the spectral densities from adjacent bins are combined by convolving the combined spectrum with the expected axion lineshape.  The grandspectrum, which is optimally sensitive to an axion, is produced by normalizing each bin of the previously generated spectrum by its total SNR such that the power fluctuation in each bin is standard Gaussian distributed with $\mu$=0 and $\sigma$=1.

Frequencies in the resulting spectrum at which the observed power exceeds a chosen threshold are then recorded.  The threshold chosen for this analysis was \SNRThreshold, which for a target significance of \SNRTarget would correspond to a \SI{10}{\percent} two-scan false negative rate in a frequentist analysis.  While the frequentist framework is no longer used, the threshold is chosen to match previous analyses for convenience.  These candidates can be the result of real RF noise sources as well as statistical fluctuations in the noise.  For each such candidate an additional observation or rescan is made to determine if the excess persists and the resulting data is processed and analyzed following the same procedure as for the initial data but with a higher order SG filter (width=\SI{30}{\kHz}, order=6) to account for the lack of tuning, as described in Ref.~\cite{brubaker2017analysis}, which gives a slightly larger attenuation of 0.76 to the expected axion signal.  Any candidate which persists above threshold in the rescan can be further interrogated with the magnetic field at \SI{0}{T}.  Because axions require a magnetic field to convert into a detectable signal, this would result in any real axion induced signal vanishing with any remaining signals resulting from RF noise from a different origin.  Signals which passed this final test would then be manually studied in follow up scans to determine if the signal has the necessary characteristics of an axion, such as the spectral shape, scaling with field and scaling with cavity profile.

\section{Axion Search Results}
\label{sec:results}

\begin{figure*}
    \centering
    \includegraphics[width=\textwidth]{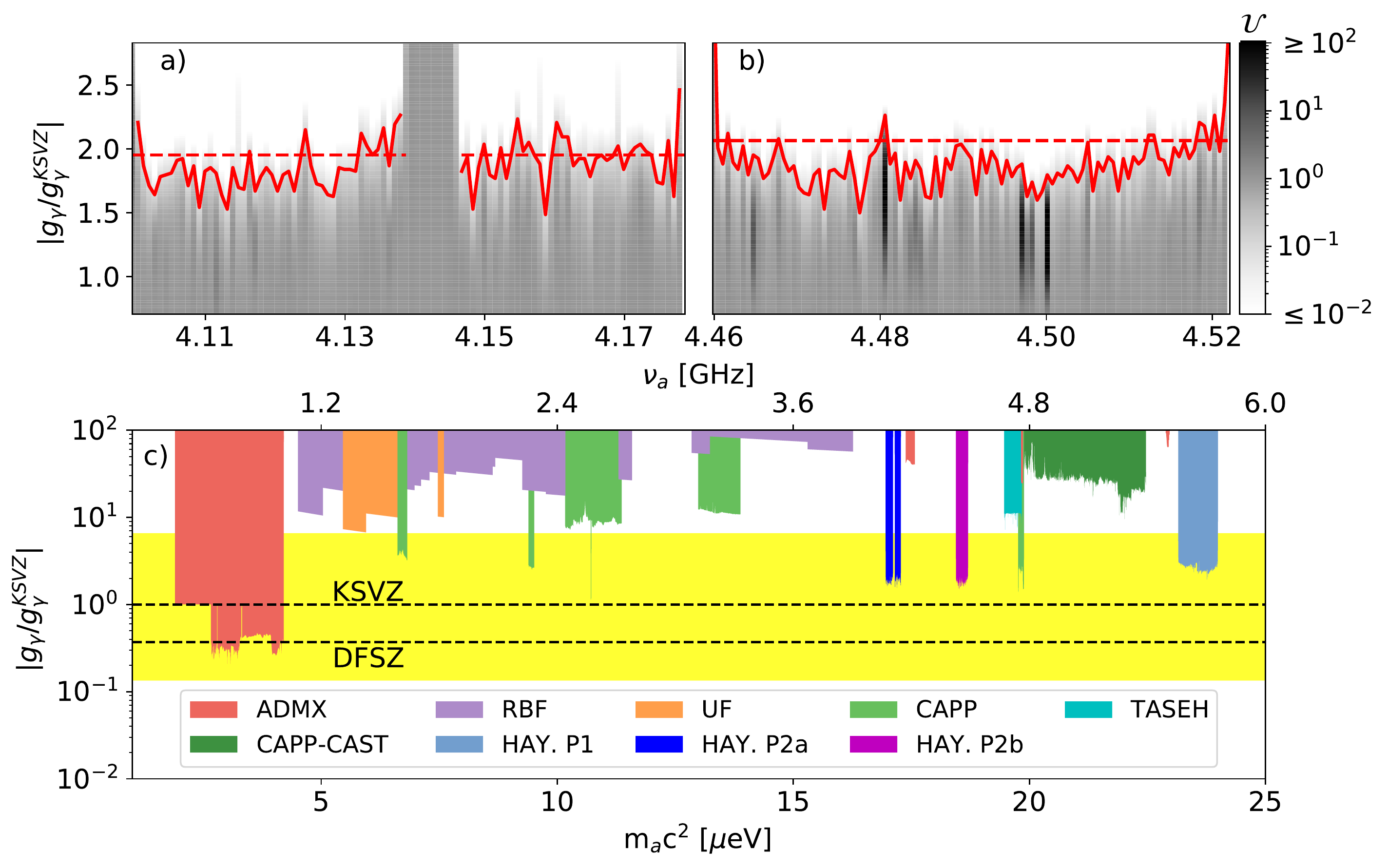}
    \caption{Axion exclusions showing the two-dimensional prior update $\mathcal{U}$ in greyscale (top panel) for (a) \PIIa \cite{backes2021quantum} and (b) \PIIb .  This includes the \SI{10}{\percent} prior update contour (solid red) as well as the  \SI{90}{\percent} aggregate exclusion (dashed red). (c) Results of this work are shown alongside previous exclusion results from HAYSTAC \PI \cite{brubaker2017first,zhong2018results} as well as from other haloscopes RBF~\cite{RBF_1987dk}, UF~\cite{uf_1990tj} ADMX~\cite{ADMXsidecar_2018ogs,ADMX_2020hay,ADMX:2021mio,bartram2021search}, CAPP~\cite{Lee:2022mnc,Kim:2022hmg,Jeong:2020cwz,CAPP:2020utb,Lee:2020cfj}, CAST-CAPP~\cite{CAPPCAST_2022rtw} and TASEH~\cite{TASEH:2022vvu}. The QCD axion model band representing the most natural KSVZ and DFSZ models  are shown in yellow~\cite{Diluzio2020modelband}, with the benchmark KSVZ~\cite{kim1979KSVZ, shifman1980KSVZ2} and DFSZ~\cite{dine1981DFSZ, zhit1980DFSZ2} model lines shown as black dashed lines.}
    \label{fig:exclusion_plot}
\end{figure*}

Data taken during \PIIa was previously analyzed to search for the presence of axion dark matter and the results were published in in Ref.~\cite{backes2021quantum} and are shown in Fig.~\ref{fig:exclusion_plot}.  No signal was observed in the mass range \PIIamrange resulting in a aggregate exclusion over this region of $g_{\gamma} \geq$ \PIIaBayAgg at the \SI{90}{\percent} level.  This result differs by a factor of $\sqrt{2}$ from that originally published in \cite{backes2021quantum}.  The correction was needed to properly handle the noise in the IF, which previously incorrectly assumed a $1/4$ quantum of noise as the quantum-limited value of the input referred noise in the IF spectrum rather than the correct value of $1/2$.  This results in the noise being underestimated by exactly a factor of 2.  The analysis has since been corrected to properly treat the IF noise and the results presented here for both \PIIa and \PIIb reflect the correct analysis. Below are details and results from the analysis of data taken during \PIIb.

Initial data was taken over a $\sim$53 day period between July 1 and August 20, 2021 covering a \PIIbfwidth region between \PIIbfrange and consisting of 801 total spectra spaced $\sim$\SI{80}{\kHz} apart.  Using the processing routine described in Sec.~\ref{sec:analysis}, these spectra are optimally summed to produce a single grandspectrum showing the observed power at each probed frequency and axion candidates are identified with a \SNRThreshold threshold.   Using two independent analyses of the data, 37 potential axion candidates were identified with 19 of the candidates being common to both. Each of the remaining candidates were found to be near threshold in both, with small differences in the analyses causing the discrepancy.  While many of these candidates are likely the result of statistical fluctuations in the noise, both parts of \PII data taking also observed a set of non-axionic RF noise candidates which presented as large $\sim$\SI{40}{\kHz} wide power deficits in the grandspectrum.  This unique structure made these candidates easy to identify, resulting in 5 candidates in \PIIa and 1 in \PIIb.  Because these regions have poor sensitivity to an axion, an RF cut was applied to remove data in the \SI{200}{\kHz} region around each candidate in the initial scans.  The source of this noise is currently unknown but so far it has been observed to vanish during the rescans, indicating this noise is either time dependent or related to the different configuration used during rescans.  This allowed for the resulting gaps in data to be filled with additional observations at these frequencies taken during rescans.      

Rescans of each of the observed power excesses were performed between September 20 and November 18, 2021.  Because this data is taken with the intention of probing individual frequencies, rather than multiple frequencies simultaneously, the optimal search configuration is slightly different than for the initial scans.  Instead, the cavity is operated near critical coupling ($\beta=1$) in order to maximize the peak sensitivity at the frequency of interest with the squeezer turned off. Analysis of the rescan spectrum found that no excess persisted above the target threshold, ruling out all candidates as potential axion signals.

While no candidate was observed to persist, an exclusion on \ggam over the mass range scanned can be set by applying the Bayesian framework outlined in Ref.~\cite{palken2020improved}. This framework is a straightforward application of Bayes' theorem and gives the change in probability of the axion existing at each frequency which was probed.  These prior updates ($\mathcal{U}$) are determined by both the experimental sensitivity and actual observed power at each frequency allowing for a “maximally informative” result. By combining the prior update from both the initial data as well as the rescan data, a \SI{90}{\percent} aggregate exclusion of $g_{\gamma} \geq$ \PIIbBayAgg is found over the entire mass range \PIIbmrange.  This result is summarized in Fig.~\ref{fig:exclusion_plot}, which shows both the aggregate exclusion as well as the full two-dimensional space of $\mathcal{U}$ with the \SI{10}{\percent} prior update contour in each frequency bin.  While no candidate in \PIIb was observed to persist above threshold during rescans, a handful of candidates are shown to have large ($\geq100$) prior updates in the Bayesian framework. Each of these points in the prior update are the result of large RF noise spikes during the initial scan which are found to fall below threshold during the rescan and thus are rejected as an axion candidate.  Due to the nature of the Bayesian analysis, the prior update calculated from the observed excess in both scans can result in a large combined prior update at couplings lower than HAYSTAC’s \SI{90}{\percent} confidence limit in certain instances, such as when the initial scan has a larger then average upward fluctuation possibly due to time dependent RF noise.  The results were validated with semi-independent analyses carried out by two groups, one based at Yale and the other at Berkeley. The two analyses agreed on average to within \SI{5}{\percent} in both the aggregate exclusion and the \SI{10}{\percent} prior update contour. This level of agreement is within the \SI{8.3}{\percent} systematic uncertainty in the exclusion arising from parameter estimation as described in Appendix~\ref{sec:uncer}.  In addition, Fig.~\ref{fig:exclusion_plot} shows the results from both \PI and \PII of HAYSTAC alongside results from other axion haloscopes in the mass region 1-\SI{25}{\micro\eV}.

\section{Conclusion}

The results presented here detail the design and performance of HAYSTAC's \PII data taking campaign and include new results  from the axion dark matter search taken during \PIIb.   This expands on the previously published results from \PIIa~\cite{backes2021quantum,Kelly_thesis, palken2020thesis} and includes specific focus on the upgrades made prior to \PII to incorporate the SSR needed to achieve sub-quantum limited noise.  In addition, results from  \PIIb are presented for the first time along with a detailed outline of updates made to facilitate a search for axions above the range covered in \PIIa.  No evidence of an axion induced power excess is observed in this new region, resulting in an aggregate exclusion on the axion photon coupling at the \SI{90}{\percent} level of \PIIbBayAgg in the mass range \PIIbmrange.  Combined with the aggregate exclusion of \PIIaBayAgg from \PIIa over the axion mass range \PIIamrange, a total of \PIIabfwidth of axion masses has now been scanned using the quantum enhanced haloscope pioneered by the HAYSTAC experiment.  This demonstrates the ability to systematically operate a sub-quantum limited dark matter axion search over large swaths of parameter space.  Continued operation with the current JPAs will allow an additional $\sim$\SI{500}{\MHz} of yet unexplored parameter space to be covered with comparable sensitivity using the current cavity.  In addition, extension to higher frequencies can be facilitated with higher frequency JPAs as well as new cavity designs currently being characterized with a dedicated R\&D campaign~\cite{Simanovskaia:2020hox}.

\section*{Acknowledgements}
HAYSTAC is supported by the National Science Foundation under grant numbers PHY-1701396, PHY-1607223, PHY-1734006 and PHY-1914199 and the Heising-Simons Foundation under grants 2014-0904 and 2016-044. We thank Kyle Thatcher and Calvin Schwadron for work on the design and fabrication of the SSR mechanical components, Felix Vietmeyer for his work on the room temperature electronics, and Steven Burrows for graphical design work. We thank Vincent Bernardo and the J. W. Gibbs Professional Shop as well as Craig Miller and Dave Johnson for their assistance with fabricating the system's mechanical components. We also wish to thank the Cory Hall Machine shop at UC Berkeley and the efforts of Sergio Velazquez for fabricating several of the prototype components tested here.  We thank Dr.~Matthias Buehler of low-T Solutions for cryogenics advice. Finally, we thank the Wright laboratory for housing the experiment and providing computing and facilities support. 

K.M.B’s affiliation with The MITRE Corporation is provided for identification purposes only, and is not intended to convey or imply MITRE's concurrence with, or support for, the positions, opinions, or viewpoints expressed by the author (Approved for Public Release; Distribution Unlimited. Public Release Case Number 22-3819).

\appendix
\section{Detector Characterization and Parameter Uncertainty}
\label{sec:uncer}

To optimally combine each spectrum and determine the sensitivity to an axion with coupling strength \axcoup, the expected SNR as a function of \detune must be precisely estimated for each spectrum.  A set of characterization measurements are used to extract the relevant parameters needed to determined the expected axion conversion power and the total system noise.  Details about this characterization procedure, along with estimates of the uncertainty associated with each extracted parameter are presented below.      

\subsection{Characterization of Expected Signal}
\label{sec:sig_uncer}

The power expected from an axion induced signal comes from Eq.~\ref{eq:axion_power}
\begin{equation}
    \begin{split}
    \mbox{\axpwr}(\mbox{\detune}) & = \frac{1}{2}\left(\frac{g^{2}_{a\gamma\gamma}}{m^{2}_{a}} \frac{2\pi \rho_{a} \hbar^{3} c^{3}}{ \mu_{0}} \right) \left( B^{2}_{0} V C_{010}   \right) \\
        & \times \left( \nu_{c} Q_{L}  \frac{ \beta}{1+\beta} \frac{1}{1 +(2\mbox{\detune}/\Delta\nu_{c})^2} \right)
        \label{eq:axion_power_appendix}
    \end{split}
\end{equation}
reorganized here into three groups of parameters by parentheses, with a factor of $1/2$ added to account for the SSR only measuring the signal in a single quadrature. The parameters in the first set of parentheses capture the dependence on both fundamental constants as well as astrophysical parameters of axion cold dark matter. 
To be consistent with literature and other axion dark matter searches,  the axion density of $\rho_{a}$=\DarkMatterDen is used.   The axion coupling is parameterized by the dimensionless coupling $g_{\gamma}$ related to  \axcoup by 
\begin{equation}
    \mbox{\axcoup} = g_{\gamma} \frac{m_a\alpha_{e}}{\pi \Lambda^{2}}
\end{equation}
where $\alpha_{e}$ is the fine structure constant and $\Lambda$ is a parameter which encodes the dependence of the axion mass on hadronic physics and is taken to be \SI{77.6}{\MeV}~\cite{olive_review_2016}.  For KSVZ axions $|g_{\gamma}|$=0.97 ~\cite{kim1979KSVZ, shifman1980KSVZ2} while for a DFSZ axion  $|g_{\gamma}|$=0.36~\cite{dine1981DFSZ, zhit1980DFSZ2}.  Because these parameters are generic to all axion experiments and are not experimentally measured here, they are not considered in the estimate of the total uncertainty.

\begin{figure}
    \centering 
    \includegraphics[width=\columnwidth]{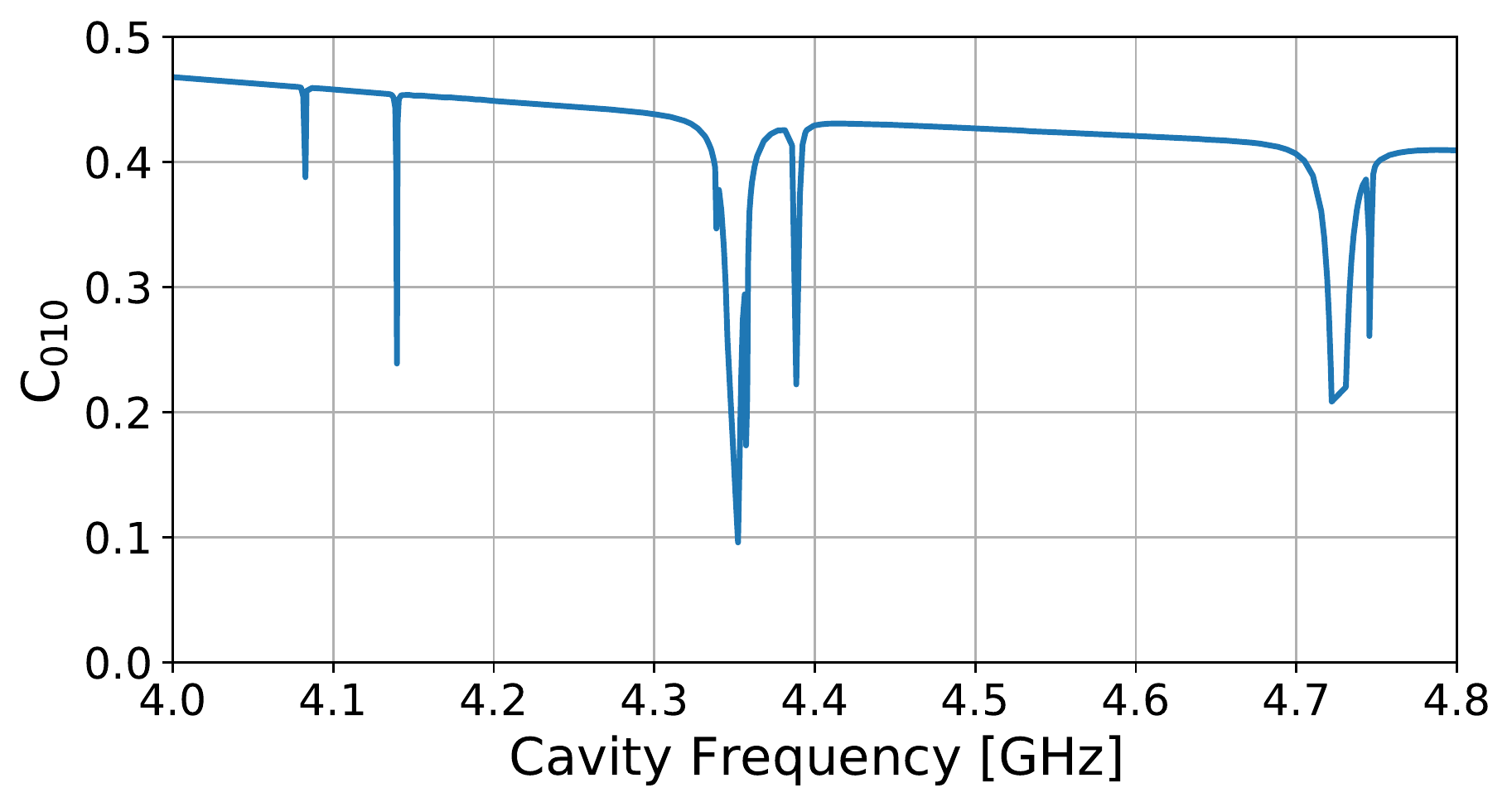}
      \caption{Plot of the cavity's form factor for the $TM_{010}$ mode over the band of interest in this paper.  The regions with sharp drops in $C_{010}$ are due to mode crossings and are avoided during the axion search.}
    \label{fig:form_factor}
\end{figure}

Contained in the second set of parentheses of Eq.~\ref{eq:axion_power_appendix} are system parameters which are assumed static over the course of a run and determined outside of the normal characterization measurements. The first of these is the the volume of the cavity, V, excluding regions occupied by the tuning rod and estimated as \cavaxionvolumeNoErr from a geometric model. Uncertainty on the volume comes from the $\sim$\SI{125}{\micro\meter} machining precision required for the alignment of the tuning rod, which approximately translates to a $\sim$0.3$\%$ uncertainty.  Second, the magnetic field, B, is set to \magfield, as measured by the superconducting current in the solenoidal magnet.  Operating the magnet in persistent mode allows for long term stability and the dominant uncertainty comes from the calibration of the current readout estimated with a Hall probe to be $\leq 1\%$ by the manufacturer \cite{private-communication}. Finally,  $C_{010}$ is calculated by modeling the spatial profile of the TM$_{010}$ electric field as a function of the rod angle with the use of CST Microwave Studio~\cite{noauthor_dassault_nodate}.  Results of these simulations showing  $C_{010}$ as a function of the TM$_{010}$ mode frequency can be found in Fig.~\ref{fig:form_factor}.  Extensive validation of this model has been summarized in Ref.~\cite{rapidis_characterization_2019} showing good agreement between the model and data.  In addition, these validations show that tilt of the tuning rod up to \SI{125}{\micro\meter} can cause $C_{010}$ to vary by $\leq 2.6\%$ relative to the baseline configuration.  Because the actual misalignment is unknown this is taken as a conservative estimate on the uncertainty.

The last set of parentheses in Eq.~\ref{eq:axion_power_appendix} contains parameters defining the cavity response which are subject to change over the course of a run and must be empirically measured each time the tuning rod is adjusted.  As described in Sec~\ref{sec:cavity} this is achieved with a pair of VNA measurements used to measure $Q_{L}$, $\nu_{c}$ and $\beta$.  The uncertainty in each of these measurements derives from two main sources.  First,  residual vibrations and other instabilities in the tuning system can cause the cavity mode to drift over the course of a single measurement which can cause time dependent fluctuations in each of these parameters.  This is most important for $\nu_\text{c}$, which needs to be held near the JPA pump frequency in order to maintain high squeezing.  In order to limit drift to within $\Delta\nu_{c}\sim\SI{60}{\kHz}$, conservatively used as a data quality cut on drift, data taking is paused after each tuning step to allow the system time to settle as described in Sec~\ref{sec:cavity}.  The residual drift is then measured by taking VNA measurements both before and after each tuning step.  The average drift over the course of \PIIa and \PIIb is measured to be \SI{18.5}{kHz} and \SI{1.5}{kHz} respectively.  This same method is used to measure the average variation in $Q_{L}$ and $\beta$ due to cavity drift, with average uncertainty on $ Q_{L}$($\beta$) found to be 0.4$\%$(0.3$\%$) in \PIIa and 0.6$\%$(0.3$\%$) in \PIIb.  In addition to drift, uncertainty in these fits can arise from bias introduced by an incomplete or inaccurate fit model.  A study using different fit windows and background models for Eq.~\ref{eq:lorenz} found an average variation of \SI{6}{\kHz} on $\nu_{c}$,  1.9$\%$ on $Q_{L}$ and 2.9$\%$ on $\beta$. Because the cavity operates at GHz frequencies the absolute uncertainty on $\nu_\text{c}$ is small but can impact the estimate on \axpwr by distorting the expected shape.  Given the observed variations in $\nu_{c}$ described above, the impact on \axpwr was estimated to be $\leq 1\%$ in both \PIIa and \PIIb.

\subsection{System Noise Calibration}
\label{sec:system_noise}

\begin{figure}[!ht]
    \centering 
    \includegraphics[trim={0cm 1cm 0.2cm 0cm},clip, width=\columnwidth]{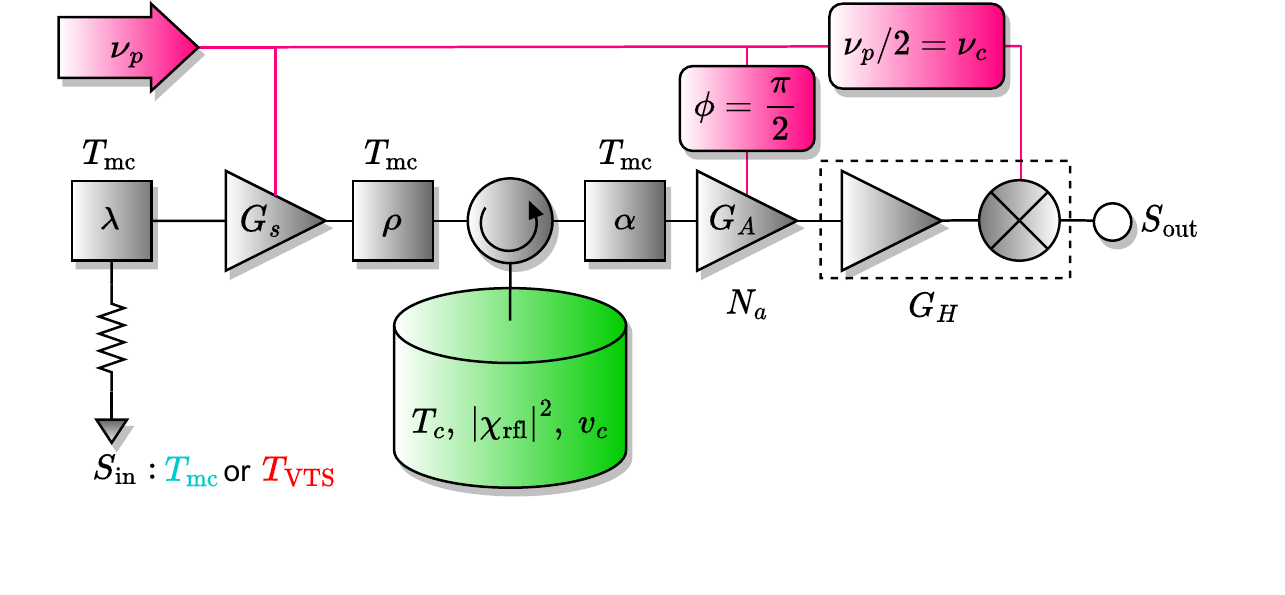}
      \caption{Model of the HAYSTAC detector used for calibration measurements described in the text.  Noise is sourced from a \SI{50}{\ohm} load which can be switched between a cold~(\TMC) or hot~($T_\text{VTS}$) load.  This noise is transmitted through a series of discrete components including loss elements ($\alpha$, $\rho$ and $\lambda$), gain elements ($G_{S}$, $G_{A}$ and $G_{H}$) and the cavity which adds noise proportional to \chirfl at its effective temperature $T_{c}$. The signal is then mixed down into the IF by half the JPA pump frequency ($\nu_p$), equivalent to the cavity's frequency ($\nu_{c}$).}
    \label{fig:det_model}
\end{figure}

In order to determine the total system noise an approximate model of the system, shown in Fig.~\ref{fig:det_model}, is used to express the expected noise spectral density coming from each part of the system.  This model is designed to capture the main components of the system without introducing unnecessary complexity.  The input of this model is sourced from the Johnson-Nyquist noise of a \SI{50}{\ohm} load held at a temperature $T$ which has noise spectral density given by
\begin{equation}
    \begin{split}
        S(T) &= \frac{1}{2} \left( \frac{1}{2} + \frac{1}{e^{h\nu/k_{B}T} - 1} \right)
        \label{eq:john_nyq_noise}
    \end{split}
\end{equation}
where the overall factor of $1/2$ in the first line accounts for measuring the noise along just a single quadrature. The temperature of this load can be switched between a cold-load held at the mixing chamber's base temperature \TMC or a hot-load held at an elevated temperature $T_\text{VTS}$ by the VTS described in Sec.~\ref{sec:cavity}.  Noise produced by either load is routed through the remaining components, which perform discrete operations on the spectral density. The multiplicative components of the model are the frequency-dependent gains $G_{S}$, $G_{A}$ and $G_{H}$ and the fractional transmissivities $\lambda$, $\rho$ and $\alpha$.  Each of these loss elements results in a fraction of the state being replaced by Johnoson-Nyquist noise at the physical temperature of the component~\cite{callen_irreversibility_1951}, all of which are assumed to be thermalized with the mixing chamber and frequency independent.   The cavity is modeled as an additional frequency dependent loss defined by its susceptibility $|\chi_\text{rfl}|^{2}$, given by the cavities reflection profile measured with the VNA as in Sec.~\ref{sec:cavity}. This results in a noise contribution proportional to its apparent physical temperature $T_{c}$,  which as described in Sec~\ref{sec:cavity} is observed to be at an elevated temperature relative to \TMC. The final gain stage encompasses gain from the HEMT as well as all subsequent stages, which includes the gain from the mixer used to mix the signal down into the IF with a LO pumped at half the JPA pump frequency, $\nu_{p}$, equivalent to the cavity frequency $\nu_{c}$.  Additional noise added by the amplification chain, also assumed to be frequency dependent, is represented by \Nadd. For an ideal parametric amplifier this noise would vanish but is in general non-zero due to imperfections causing deviations from this ideal behavior.
 
Cascading all of these elements together allows for the expected output spectral density in the IF to be expressed as 
\begin{equation}
      \begin{split}
         S_\text{out}(|\mbox{\detune}|) &=   G_{A}G_{H}   \bigg( \mbox{\Nadd} + 2\bigg[ \alpha(1 - \rho) \left| \chi_\text{rfl} \right|^{2} S_\text{mc}  \bigg.\\
          &  + \rho\alpha G_{S} \left| \chi_\text{rfl} \right|^{2} \left((1-\lambda)S_\text{mc} + \lambda S_\text{in} \right) \bigg]  \\
          & + 2\bigg.  \bigg[ \alpha\left(1 - \left| \chi_\text{rfl} \right|^{2}\right)S_{c}   + \left(1 - \alpha\right)S_\text{mc} \bigg] \bigg) %\\
         % & =  G_{A}G_{H}  \bigg( \mbox{\Nadd} + 2 N_{c} + 2 N_{r} \bigg) 
       \end{split}
       \label{eq:det_model}
\end{equation}
where the explicit frequency dependence of each variable has been suppressed and $S_{mc}$, $S_{c}$ and $S_{in}$ are the noise spectral densities of the mixing chamber, cavity and input load respectively as given by Eq.~\ref{eq:john_nyq_noise}. The terms in Eq.~\ref{eq:det_model} can be arranged to resemble the noise components in Eq.~\ref{eq:noise_components}, but now describe the variance of the $\hat{X}$ quadrature rather than its phasor elements. In this case the first term in brackets comes from the noise reflected off of the cavity, $N_{r}$, and the second term in brackets comes from the excess noise of the cavity, $N_{c}$, allowing $S_\text{out}$ to be written as
\begin{equation}
    S_\text{out} = G_{A}G_{H}  \bigg( \mbox{\Nadd} + 2 \left( N_{r} + N_{c}\right) \bigg)
\end{equation}
with the terms in parentheses being \Nsys in the IF. The factor of 2 which appears in front of $N_{c}$ and $N_{r}$ represents the noise present in the IF spectrum from positive and negative detunings in the RF,
\begin{equation}
    S^{IF}_{\text{out}}(|\mbox{\detune}|) = S^{RF}_{\text{out}}(+\mbox{\detune}) + S^{RF}_{\text{out}}(-\mbox{\detune}),
\end{equation}
added prior to the addition of \Nadd. Because both noise sources are symmetric about $\nu_{p}$, as illustrated in Fig.~\ref{fig:analysis_band}, the contribution of both is exactly doubled in each IF bin at the output of the JPA.  As a result, the quantum limited noise in one quadrature for any frequency bin in the IF is $S_{\text{out}}=1/2$ when operating at \SI{0}{\kelvin} without squeezing. Calibration of the total noise requires measuring each of the parameters in Eq.~\ref{eq:det_model}.

This procedure starts with a measurement of the three transmissivities, performed prior to the start of each run and assumed constant over the course of data taking.  The first measurement follows the procedure outlined in Ref.~\cite{malnou2019squeezed} by which the total transmission efficiency between the squeezer and amplifier ($\eta$) is extracted through an observation of the change in noise power when swapping the pre-amplification between the amplifier and squeezer while detuned from the cavity.  This procedure results in a measured transmission efficiency of $0.63\pm0.03(0.60\pm0.03)$ in \PIIa(\PIIb).  These measurements agree with the observed squeezing in the system, which is fundamentally limited by this total loss as described by Eq.~\ref{eq:squeezing_limit}.  Following this, the loss before the squeezer is extracted through a measurement of the total system transmission efficiency ($\zeta=\alpha\rho\lambda$) and dividing out the already determined $\eta$. This measurement is taken by observing the dependence of noise power on both the input load and mixing chamber temperatures, the slope of which gives the total system transmissivity observed to be $\zeta=0.50 \pm 0.07$.  The uncertainty on these two measurements come from the statistical variations over repeated measurements.  Unfortunately, the combinations of measurements do not distinguish between losses before and after the cavity and instead it is approximately assumed that these losses are evenly divided between the two components giving $\alpha=\rho=\sqrt{\eta}$.   Because $\rho$ only appears in Eq.~\ref{eq:det_model} as the total transmissivity $\eta$, this assumption only has an impact on $S_\text{out}$ through $\alpha$. An estimate on this uncertainty can be made by comparing previous measurements of $\alpha$ from \PI~\cite{Kenany2017design} which show a $\sim$\SI{10}{\percent} relative change in the transmission efficiency  for \PII.  While this drop is likely due to the addition of the more lossy triple junction circulator, a conservative estimate of $\sim$\SI{10}{\percent} is associated with $\alpha$ to account for this assumption.

\begin{table}[h!]
    \centering
    \begin{tabular}{lccccc} % used to be c||ccccc, can change back
     %\hline \hline
     Calib ID \quad & SQZ \quad & ($\nu_{p}/2$ -$\nu_{c}$) \quad & Load Temp. \quad & $\tau_{cal}$ \quad & $G_{A}$\\ 
     %ID & State &  [MHz] & Load Temp. & [s] \\
     \hline \hline
     %Afghanistan   & AF    &AFG&   004 \\
     g &      ON  &  \SI{0}{\MHz} & \TMC &  \SI{10}{\second} & \multirow{2}{*}{$G^{g}_{A}$} \\
     b &      OFF &  \SI{0}{\MHz} & \TMC & \SI{10}{\second} &  \\
     \hline
     a &      OFF &  -\SI{3}{\MHz} & \TMC & \SI{10}{\second} & \multirow{2}{*}{$G^{c}_{A}$} \\
     cold(c) &   OFF &  -\SI{3}{\MHz} & \TMC & \SI{900}{\second} & \\
     \hline 
     hot(h) &    OFF &  -\SI{3}{\MHz} & $T_\text{VTS}$ & \SI{900}{\second} & $G^{h}_{A}$\\
     %\hline \hline
    \end{tabular}
    \caption{Summary of the different configurations of the system during the periodic calibration measurements used for the calculation of \Nadd, $G_{s}$ and $S_{c}$. }
    \label{tab:calib_summary}
\end{table}

\begin{figure}
    \centering 
    \includegraphics[width=\columnwidth]{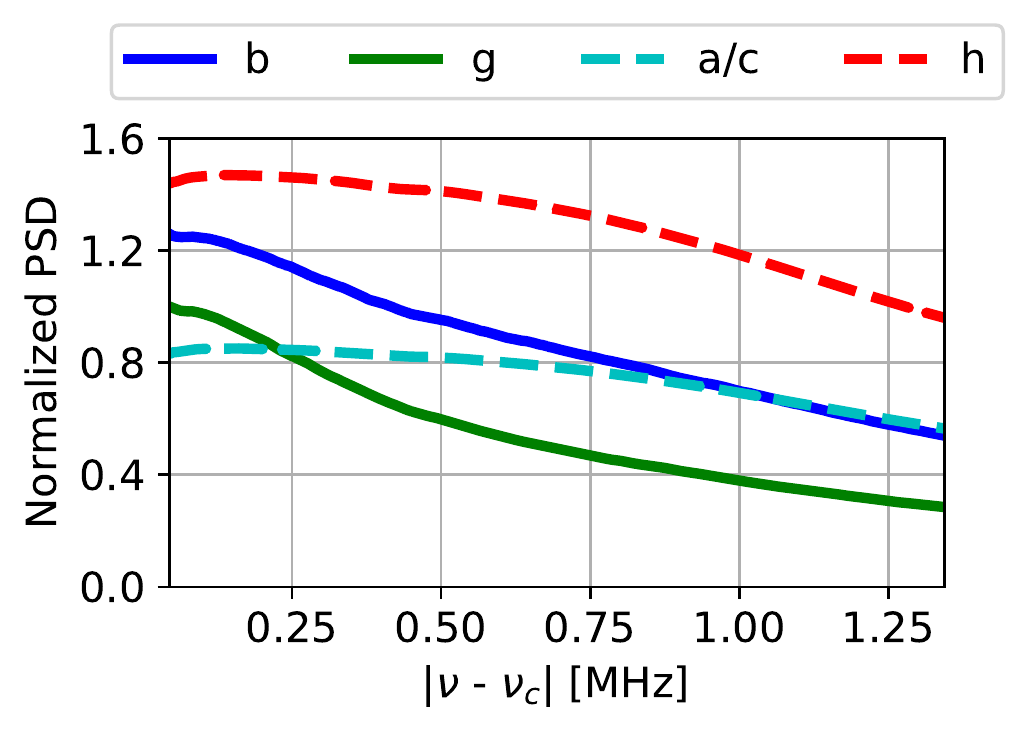}
      \caption{Example set of calibration spectra taken periodically to extract parameters for determining the system noise, shown for both on resonant (solid lines) and off resonant (dashed lines) measurements. The spectra are smoothed to remove expected statistical fluctuations and normalized relative to the on resonant power of the "g" measurement, which is taken in the same configuration as that used to search for axions. }
    \label{fig:abg_spec}
\end{figure}

Extraction of the remaining parameters is done by periodically collecting a set of five power spectrum measurements taken with the DAQ described in Sec.~\ref{sec:DAQ} to record the average power spectrum over an integration time $\tau$.  Each measurement is taken with certain components removed, allowing for the dependence of each parameter to be isolated with a comparison of the relative change in the spectral shape defined by the ratio 
\begin{equation}
    Y^{ij} = \frac{S^{i}_{\text{out}}}{S^{j}_{\text{out}}}\frac{G_{A}^{j}}{G_{A}^{i}}
    \label{eq:yfac}
\end{equation}
were $i$ and $j$ represent the calibration measurement ID defined in Table~\ref{tab:calib_summary}.  While each measurement is short relative to the integration time the axion sensitive noise measurement taken at each tuning step, the total time required for the full set of measurements is $\sim$\SI{1}{hr} and is largely dominated by time spent waiting for the system to reach equilibrium after switching between the cold and hot load.  For this reason, these measurements are roughly taken once every nine spectra so as to minimally impact the deadtime discussed in Sec.~\ref{sec:DAQ}.  An example set of power spectrum taken during the calibration routine are shown in Fig.~\ref{fig:abg_spec} where the spectrum are normalized to the on resonant power of the "g" measurement.

In order to limit uncertainty in the parameter extraction, each measurement would ideally be taken such that only the intentionally modified parameters would change between measurements.  While this is possible for most parameters in the model, it is not possible for the amplifier gain which is subject to change when re-biased as is required for the off-resonant and hot-load measurements.  Direct measurements of the single quadrature gain are made each time the amplifier bias is changed but these measurements have large uncertainties which can result in deviations of the calibration parameters outside of their physically realistic regions if not corrected.  Because the gains enter the calculation in Eq.~\ref{eq:yfac} only  as ratios, realistic estimates of each parameter are extracted by allowing the gain ratios $G^{bg}_{A}/G^{ac}_{A}$ and $G^{ac}_{A}/G^{h}_{A}$ to deviate from their approximately measured value.  In order to avoid overestimating the detector sensitivity, the set of gains which produce physically realisable values are scanned and the pair which give the highest total noise are conservatively used as the final estimate.         

The first two measurements are taken with $S_\text{in}$ sourced by a load at \TMC and the amplifier tuned to match the cavity frequency, the same as for axion data. Isolation of $G_{S}$ is achieved by switching the JPA used to prepare the squeezed state on for measurement "g" and off in measurement "b", allowing for $G_{S}$ to be found as:
\begin{equation}
    \begin{split}
    G_{S} & = \frac{1}{2\alpha\rho \left| \chi_\text{rfl} \right|^{2}Y^\text{bg}S_\text{mc} }  \biggl[ 2 N^{b}_{c}(1-Y^\text{bg}) + 2 N_{r}^{b}  \biggr. \\
     & \biggl. + \mbox{\Nadd}(1-Y^\text{bg})   - 2\alpha(1-\rho) \left| \chi_\text{rfl} \right|^{2}Y^\text{bg} S_\text{mc}  \biggr]
    \label{eq:gs_func}
    \end{split}
\end{equation}
The parameter scan restricts values of $G_{S}$ to the range $0<G_{S}<1$, where the upper bound asserts that squeezing corresponds to gain less than 1 and the lower bound is a physical requirement.  As can be seen in Eq.~\ref{eq:gs_func}, both $S_{c}$ and \Nadd still appear in the extraction of $G_{S}$.  As a result, the uncertainty on $G_{S}$ is relatively high and an estimate of its value is derived by comparing the extracted $G_{S}$ to the approximate target gain of \SI{13}{\dB} needed to achieve optimal squeezing.  This results in a $\sim$\SI{50}{\percent} uncertainty on the extraction of $G_{S}$ in both phases.  While this uncertainty is high, the relative contribution to the total noise and in turn the exclusion is small.   

Following this a measurement, "a", is taken with the squeezer still off but the amplifier tuned 3~\si{MHz} below the cavity's resonant frequency. Comparing this to measurement "b", the cavity's excess temperature is found as
\begin{equation}
    S_{c} = \frac{\left(Y^\text{ba} - 1\right)\left(\mbox{\Nadd} + 2S_{mc}\right) + 2\alpha S_{mc}\left( 1 - \left| \chi_\text{rfl} \right|^{2}\right)}{2\alpha(1 - \left| \chi_\text{rfl} \right|^{2})}
    \label{eq:SC_full}
\end{equation}
Bounds on $S_{C}$ are given by an absolute measurement of the cavity's physical temperature taken at the start of the run.  This is performed by raising the mixing chamber temperature until the excess cavity noise vanished at $\sim$\SI{225}{m\kelvin}, indicating the system was in thermal equilibrium.  While this would imply $S_{c}\approx0.6$, measurements of the cavity's exterior temperature indicate that only an internal component is at an elevated temperature. Currently this is assumed to be the tuning rod, which as described in Sec.~\ref{sec:cavity} has only a weak thermal link to the mixing chamber.  As a result it is assumed that the effective cavity noise temperature $T_{C}<T_{\text{rod}}$ and a bound of $0.38<S_{c}<0.44$ is derived from the rod's geometric orientation relative to the antenna.  In addition, an approximate uncertainty of \SI{20}{\percent} is assigned to $S_{C}$ based on the initial estimate of $T_{\text{rod}}$ taken before the start of the run.        

Finally, extraction of \Nadd is performed with a pair of measurements where $S_{in}$ is switched between a cold load (\TMC) and a hot load ($T_\text{VTS}$)  with the amplifier tuned \SI{3}{\MHz} below the cavity's resonant frequency and the squeezer switched off.  This technique is similar to the Y-Factor measurement presented in Ref.~\cite{Kenany2017design,brubaker2017first} but makes use of the VTS which allows for control of the load temperature below \SI{775}{m\kelvin}. The ratio of the two measurements then gives \Nadd as
\begin{equation}
    \mbox{\Nadd} = \frac{2\left(N_{c}^{h} + N_{r}^{h} - Y^{hc}\left(N_{c}^{c} + N_{r}^{c}\right)\right)}{Y^\text{hc} - 1}
    \label{eq:NA_full}
\end{equation}
Because neither the calculation of $S_{c}$ nor \Nadd are fully decoupled from each other, their calculation is done in two iterations.  First, \Nadd is calculated by ignoring the cavity's contribution to Eq.~\ref{eq:NA_full}. The resulting estimate is used in Eq.~\ref{eq:SC_full} to calculate $S_{c}$, which is then fed back to correct for the cavity's contribution to the determination of \Nadd.  A lower bound of $\mbox{\Nadd}>0.008$ is set based on the HEMT's added noise spec (LNF-LNC4 8A~\cite{HEMT_LNF}) and no upper limit is applied.  Uncertainty on \Nadd is largely due to uncertainty in the temperature of the VTS, which is known to within \SI{5}{m\kelvin} from the Magnicon temperature sensor described in Sec.~\ref{sec:cavity}, as well as uncertainty in the total transmission efficiency $\zeta$ which give a total uncertainty of $63\%$ in both phases.    

\begin{figure}
    \centering 
    \includegraphics[width=\columnwidth]{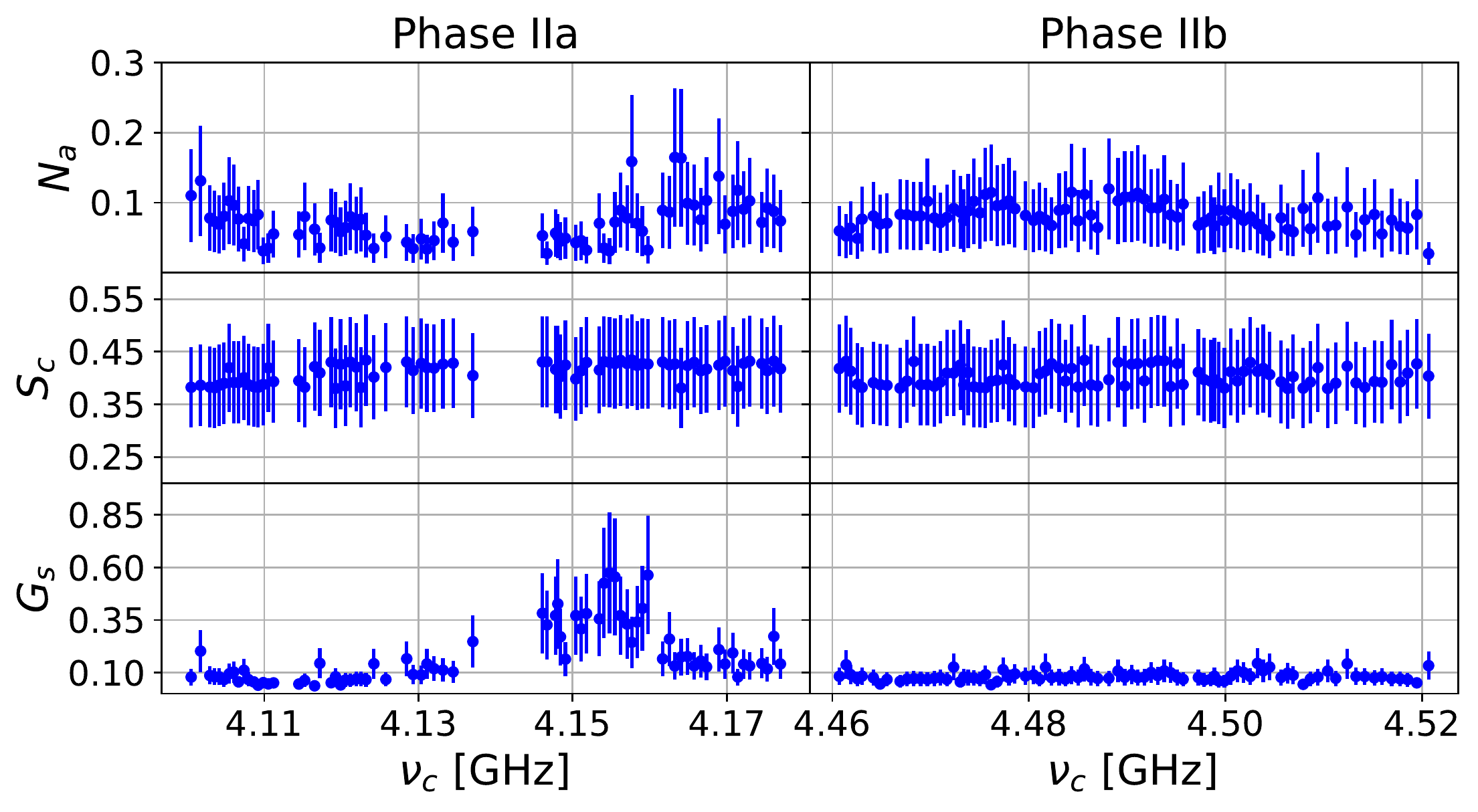}
      \caption{Average value over the analysis band of the three model parameters extracted from the periodic calibrations used to determine the system noise, shown for all of \PII as a function of the cavity frequency.  Error bars are the result of the systematic uncertainty on each parameter shown in Table~\ref{tab:exclusion_unc}.}
    \label{fig:calib_vals}
\end{figure}

\begin{figure}
    \centering 
    \includegraphics[width=\columnwidth]{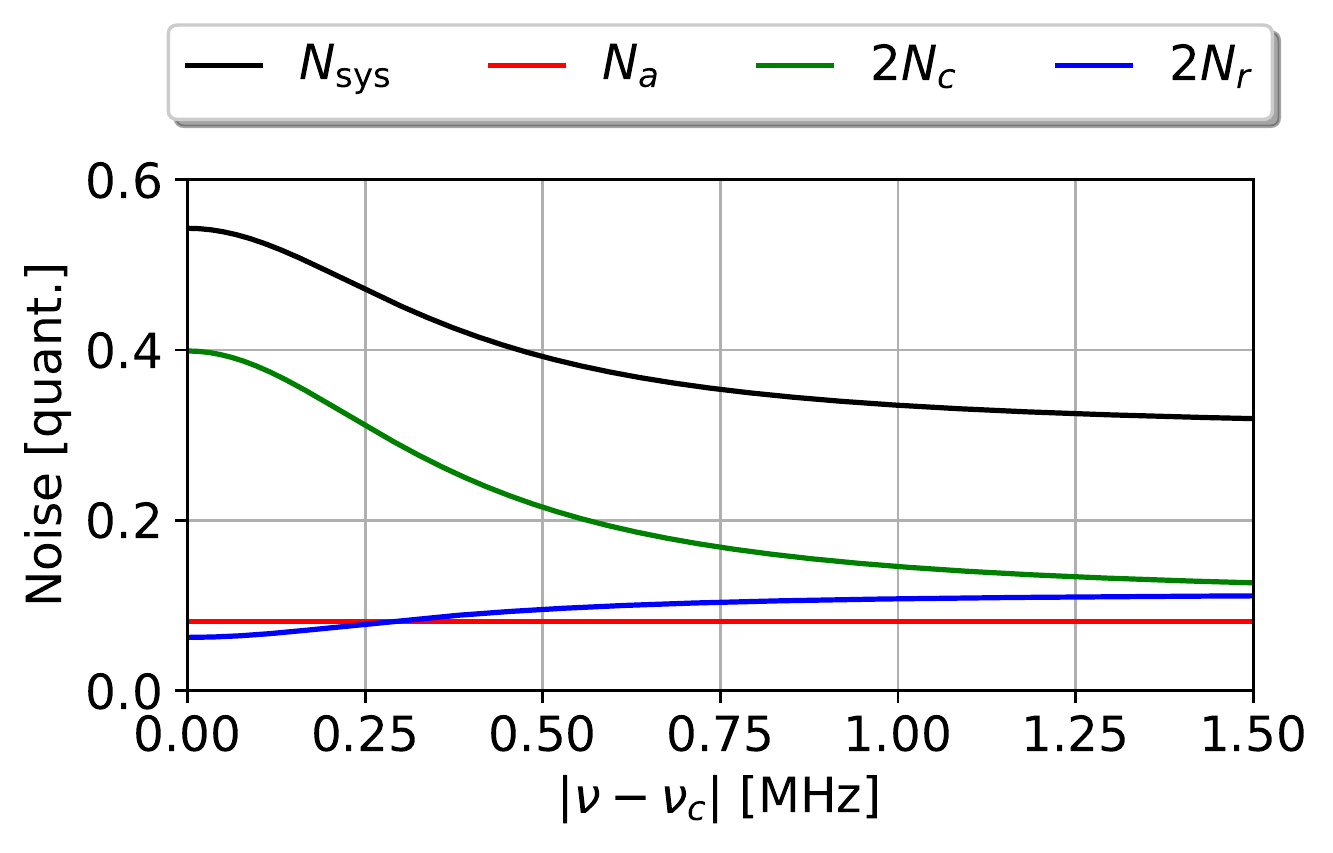}
      \caption{Contributions to \Nsys in the IF,   plotted in units of power spectral density as a function of detuning using the average calibration parameters from \PIIb shown in Table~\ref{tab:exclusion_unc}.}
    \label{fig:total_sys_noise}
\end{figure}

The extracted values for $G_{S}$, $S_{C}$ and \Nadd for each set of calibration measurements, reported as the average over the analysis band, is shown in Fig.~\ref{fig:calib_vals} for both data taking phases.  Using the extracted value for each parameter, the total system noise can be calculated as a function of detuning using Eq.~\ref{eq:det_model}, as shown in Fig.~\ref{fig:total_sys_noise} in the IF using the average parameters from \PIIb.  As expected, the SSR reduces the added noise from $N_{r}$ over a large range of \detune allowing for an improved measurement bandwidth when operating strongly overcoupled to the cavity.

\subsection{Uncertainty on Exclusion}
\label{sec:tot_uncertainty}

\begin{table}[t]
    \addvbuffer[12pt 8pt]{\begin{tabular}{c||cc|cc} 
    \hline \hline
             & \multicolumn{2}{c|}{Value}   &  \multicolumn{2}{c}{Frac. Unc.}  \\
    Par.     & \PIIa & \PIIb & $\sigma_{\text{par}}$ & $\sigma_{g_{\gamma}}$ \\ \hline
    $\nu_{c}$   &  \SI{4.14}{\GHz} &   \SI{4.49}{\GHz} & -  & $\leq 0.4\%$ \\
    $B_{0}$     &  \SI{8}{\tesla} &  \SI{8}{\tesla} & $1\%$ & 1$\%$  \\
    $V$         &  \cavaxionvolumeNoErr &  \cavaxionvolumeNoErr & 0.3$\%$  & $0.15$\% \\
    $C_{010}$   &  0.43 &  0.43 & $2.6\%$  & 1.3$\%$ \\
    $Q_{L}$     &  6180 & 5915 & 2.0$\%$  & 0.6$\%$   \\
    $\beta$     &  6.9  & 6.5  & 2.9$\%$  &  0.1$\%$ \\
    \hline
    $\alpha\rho$ &   0.63  & 0.60  & $2\%$ & $1\%$ \\
    $\alpha$     &   0.794 & 0.775 & $10\%$ & $4.5\%$  \\
    
    \hline
    $G_{S}$   &   0.11 & 0.08 & $50\%$ & $1.5\%$   \\
    $S_{C}$   &   0.41 & 0.40 & $20\%$ & $3.5\%$  \\
    \Nadd &   0.08 & 0.08 & $63\%$ & $5.5\%$    \\
    
    \hline \hline %\hline 
     \rule{0pt}{3ex} $g_{\gamma}$   & \PIIaBayAgg & \PIIbBayAgg &  - & $8.3\%$ \\

    \end{tabular}}
    \caption{Breakdown of each experimentally determined parameter which is needed to determine the expected \SNR for an axion-induced signal.  For parameters which vary between tuning steps, the average value over the entire phase is used.  Details about the quantification of each parameter and the associated uncertainty are given in the text.}
    \label{tab:exclusion_unc}
\end{table}

As many of the parameters used to calculate \SNR vary with \detune, their relative contribution to the total uncertainty also vary.  To account for this, the total \SNR is calculated by generating a set of fake spectra matching the frequencies in real data. For each spectrum, the individual \SNR is calculated using the average parameters from the phase and shifted by the estimated uncertainty on each parameter.  The total \SNR, resulting from the squared sum of overlapping spectra,  is calculated and used to estimate the overall bias in the exclusion on $g_{\gamma}$.  The results of this study are summarized in Table~\ref{tab:exclusion_unc}, which also includes a summary of the individual uncertainty contribution from each parameter. The total uncertainty on $g_{\gamma}$ is \SI{8.3}{\percent} with the largest contributions coming from uncertainty in $S_{C}$, $\alpha$ and $N_{A}$.  Currently this uncertainty is not propagated to final exclusion and is instead included here as an approximate estimate on the reliability of the final exclusion.  A full treatment of this uncertainty is needed to account for correlations in the extracted parameters introduced by the calibration procedure.

\bibliographystyle{apsrev4-1} 
\bibliography{all_references}

\end{document}